\documentclass[11pt,a4paper,dvipdfmx]{article}
\selectfont

\usepackage{longtable}
\usepackage{booktabs,multirow}
\usepackage{amsmath,amsthm,latexsym,amssymb}
\usepackage{graphicx,bm,fancybox,multirow,hhline,indentfirst}
\usepackage{ascmac,atbegshi}
\usepackage{natbib,verbatim,color}
\usepackage[colorlinks,citecolor=blue,urlcolor=blue,dvipdfmx]{hyperref}
\usepackage{mleftright}

\makeatletter
\def\section{\@startsection {section}{1}{\z@}{-3.5ex plus -1ex minus -.2ex}{2.3 ex plus .2ex}{\large\sc\centering}}
\def\subsection{\@startsection {subsection}{1}{\z@}{-3.5ex plus -1ex minus -.2ex}{2.3 ex plus .2ex}{\large}}
\makeatother

\theoremstyle{definition}
\newtheorem{theorem}{Theorem}
\newtheorem{corollary}{Corollary}
\newcommand{\argsup}{\mathop\text{argsup}}

\title{\Large\bf Information criteria for detecting change-points in the Cox proportional hazards model}

\medskip

\author{\fontsize{11}{15}\selectfont
			Ryoto Ozaki\\ \fontsize{10}{15}\selectfont
			Biometrics Department, Chugai Pharmaceutical Co., Ltd.\\ \fontsize{10}{14.4}\selectfont
			Department of Statistical Science, The Graduate University for Advanced Studies
			\medskip\and \fontsize{11}{15}\selectfont
			Yoshiyuki Ninomiya\thanks{Corresponding author. 10-3 Midori-cho, Tachikawa-shi, Tokyo 190-8562, Japan. E-mail: ninomiya@ism.ac.jp}	
			\\ \fontsize{10}{15}\selectfont
			Department of Statistical Inference and Mathematics, The Institute of Statistical Mathematics	
			\\ \fontsize{10}{15}\selectfont
			Department of Statistical Science, The Graduate University for Advanced Studies}
\date{}

\allowdisplaybreaks

\def\E{\text{E}}
\def\V{\text{V}}
\def\P{\text{P}}
\def\T{\mathsf{T}}
\def\O{\text{O}}
\def\o{\text{o}}
\def\OP{\text{O}_\text{P}}
\def\oP{\text{o}_\text{P}}
\def~{\hphantom{0}}

\begin{document}

\maketitle

\begin{abstract}
The Cox model, which is commonly used for clinical trials in which the time to clinical events such as death or to the occurrence of specific adverse events is of interest, assumes proportional hazards and log-linearity. However, it has been shown that proportional hazards do not hold in cases such as the delayed onset of a treatment effect. Moreover, analyses under such deviations reduce the detection power for the effect of covariates and make it difficult to interpret the estimated hazard ratios. In such a situation, the survival curves are expected to overlap for a certain period after the start of treatment, and then the difference between the curves increases. As this is considered to be an acute change in the hazard ratio function, change-point analysis is important in survival time analysis. Hence, this paper considers the Cox proportional hazards model with change-points and derives AIC-type information criteria for detecting those change-points. Because of its irregularity, a change-point model does not allow for conventional statistical asymptotic theories; thus, using a formal AIC that penalizes twice the number of parameters would clearly cause overfitting. Accordingly, we construct specific asymptotic theories by using the partial likelihood estimation method in the Cox proportional hazards model with change-points. By applying the original AIC derivation method, we propose information criteria that are mathematically guaranteed. If the partial likelihood is not used in estimation, then asymptotic theories of change-point analysis may not necessarily be applicable, whereas if the partial likelihood is used, then information criteria with penalties much larger than twice the number of parameters can be obtained explicitly. Numerical experiments confirm that the proposed criterion, in comparison to the formal AIC, more accurately approximates the asymptotic bias to the risk function and is clearly superior in terms of the AIC's original purpose of providing an estimate that is close to the true structure. We also apply the proposed criterion to actual clinical trial data to indicate that it will easily lead to different results from the formal AIC. 

\medskip

\end{abstract}

\section{Introduction}
\label{sec1}
The proportional hazards model proposed by \cite{Cox72} is widely used for survival analysis in clinical studies with time-to-event endpoints. This model involves potential assumptions such as proportional hazards and log-linearity. Proportional hazards assume that the hazard ratio is constant over time, but the data structure in actual clinical trials often deviates from this assumption. Various problems have been pointed out here: the lack of theoretical validity in applying analytical methods for which the proportional hazards property is assumed in situations where it does not hold; the reduced detection power for covariates; and the difficulty of interpreting estimated hazard ratios (see, e.g., \citealt{uno2014moving}). The problem of deviations from the proportional hazards assumption has been around for a long time, and works such as \cite{gill1987simple} and \cite{hess1995graphical} proposed methods to detect and avoid deviations from the assumption.

In clinical trials comparing therapeutic products such as immune checkpoint inhibitors or cancer vaccines with control products such as placebos, it takes time from the start of treatment to the onset of efficacy because of the products' mode of action (\citealt{fda11vac}). The survival curves in such clinical trials are expected to overlap for a certain period after the start of treatment, and then the difference between the curves increases. Because this is considered to be an acute change in the hazard ratio function, change-point analysis is important for detecting such time points in survival time analysis. \cite{liu2008monte} and \cite{he2013sequential} proposed methods based on the maximum score test and a sequential test approach, respectively, to detect change-points in a hazard ratio that is assumed to be constant in segmental terms. However, there are still no reported methods for detecting change-points by using information criteria.

Accordingly, in this paper, we use a combination of the Cox proportional hazards model and a change-point model to develop information criteria that are theoretically valid. Specifically, we use the partial likelihood and add a regularization term to the loss function used for estimation. Then, from the viewpoint of the conventional Akaike information criterion (AIC, \citealt{Aka73}), we assume the existence of $m$ change-points in developing the AIC for a model with $m$ change-points. Our derivation of asymptotically unbiased estimators for a properly defined risk function is theoretically based on \cite{Tsiatis1981large}, which derived the asymptotic properties of regression parameter estimators via the partial likelihood method, and \cite{pons2002estimation}, which evaluated asymptotic theories in a change-point model.

Because a change-point model requires specific asymptotic theories (see, e.g., \citealt{CsoHor97}), various theories have been developed for the test-based approach, which implies that the information criteria for a model with change-points also require specific theories. \cite{Sie04} first derived an information criterion based on the original definition by using theory specific to the change-point model; that is, the approach was based on asymptotically valid rather than formally. Hence, the purpose of this paper is to derive AIC-type information criteria. For a basic change-point model in which data are observed independently at each time, such a derivation was performed by \cite{Nin15}. That work, based on the original AIC definition, gave the criterion as an asymptotically unbiased estimator of the Kullback-Leibler divergence between the true and estimated distributions. According to that derivation, the asymptotic bias due to the regression parameter is 1, whereas that due to the change-point parameter is 3, i.e., three times larger. This implies that consideration of theory specific to a change-point model in the AIC derivation would significantly alter the analytical results.

The Cox proportional hazards model with change-points can be regarded as a model that assumes that proportional hazards do not hold for the entire follow-up period but do hold for segmented periods. We also assume log-linearity in each interval, as in the conventional Cox proportional hazards model. This log-linearity implies that the hazard function yields a linear expression when the logarithm of its covariate-dependent part is taken. In observational studies, it is rare to measure all the covariates that are necessary for hazard ratio estimation; moreover, even if log-linearity is established, the model used for estimation may be misspecified. Of course, there is a deviation from log-linearity when the effect of covariates cannot be expressed by a linear equation. In addition, if the regression parameters are estimated by the conventional partial likelihood method in such a case, the estimator will not have good properties such as consistency, which makes it difficult to interpret the hazard ratios. Estimation under conditions of model misspecification in the Cox proportional hazards model was discussed in \cite{struthers1986misspecified} and \cite{lin1989robust}. In this paper, by leveraging those discussions, we also extend the proposed AIC to handle such conditions of model misspecification, in the form of the Takeuchi information criterion (TIC, \citealt{Tak76}).

We note here that in the Cox proportional hazards model with change-points, unlike the usual change-point model, the time and outcome variables are not observed separately, but both are the same variable of survival time. When the time is not observed, or when both are the same but with added noise, this change-point model will have the same type of structure as the so-called mixture distribution models. As described in \cite{DacG99}, mixture distribution models also have asymptotic properties that are different from those of regular statistical models and even more different from those of change-point models. In other words, it is not at all obvious whether the AIC addressed in this paper is of the same type as the AIC for a conventional change-point model or the AIC for a mixture distribution model. The information criteria proposed in this paper should thus clarify this point.

We organized the rest of this paper as follows. In Section \ref{sec2}, for preliminaries, we first define the Cox proportional hazards model with change-points and assume the usual conditions for asymptotic theories in change-point analysis. Then, by using the original AIC derivation, we define an AIC-type information criterion as an asymptotically bias-corrected version of the regularized maximum log-partial likelihood. In Section \ref{sec3}, we evaluate the asymptotic bias and show that it can be expressed explicitly. To evaluate our approach, we describe the results of numerical experiments in Section \ref{sec4}. First, we confirm that the asymptotic bias evaluation does indeed approximate the bias accurately. Then, the performance of the derived AIC is compared with that of the formally defined AIC without using any theory specific to the change-point model. For further evaluation, in Section \ref{sec5}, we describe a case study on applying the derived AIC and the formally defined AIC to actual clinical trial data for change-point detection. In Section \ref{sec6}, we extend our theory and proposed AIC to the case where model misspecification is allowed, and we derive TIC. Finally, we give our conclusion in Section \ref{sec7}.

\section{Preliminaries}
\label{sec2}
\subsection{Model and assumptions}
\label{sec2_1}
For the Cox proportional hazards model with shifts in the regression parameters, we incorporate a model with $m$ change-points as follows:
\begin{align}
	\label{CoxRegCP}
	\lambda(t\mid \bm{z})=\lambda_0(t)\exp(\bm{\beta}^{(j)\T}\bm{z}), \qquad t\in [k^{(j-1)},k^{(j)}), \qquad j\in\{1, 2, \ldots, m+1\}.
\end{align}
Here, $\lambda(t\mid \bm{z})$ is the hazard function under a given covariate vector $\bm{z}$, and that $\lambda_0(t)$ is the baseline hazard function. For each $j\in\{1, 2, \ldots, m+1\}$, let the regression parameter $\bm{\beta}^{(j)}$ be a $p$-dimensional vector; that is, $\bm{\beta}\equiv(\bm{\beta}^{(1)\T},\bm{\beta}^{(2)\T},\ldots,\bm{\beta}^{(m+1)\T})^\T$ is a $p(m+1)$-dimensional vector. Let $\bm{k}\equiv(k^{(1)},k^{(2)},\ldots,k^{(m)})^\T$ be an $m$-dimensional vector of change-point parameters, and let $k^{(0)}=0$ and $k^{(m+1)}=T$, where $T$ is the follow-up period of the survival time. Also, suppose that the true values of $\bm{k}$ and $\bm{\beta}$ are $\bm{k}^*=(k^{*(1)},k^{*(2)},\ldots,k^{*(m)})^\T$ and $\bm{\beta}^*=(\bm{\beta}^{*(1)\T},\bm{\beta}^{*(2)\T},\ldots,\bm{\beta}^{*(m+1)\T})^\T$, respectively, such that $0<k^{*(1)}<k^{*(2)}<\cdots<k^{*(m)}<T$. To construct a change-point model, we assume that
\begin{align}
	\label{ConditionBeta}
	\bm{\beta}^{*(1)}\ne\bm{\beta}^{*(2)}\ne\cdots\ne\bm{\beta}^{*(m+1)},
\end{align}
and that $\bm{k}^*$ and $\bm{\beta}^*$ are unknown. For simplicity, let the space of $\bm{\beta}$ be compact.

The regression parameters in the Cox proportional hazards model are usually estimated by the partial likelihood method proposed by \cite{Cox72}. First, let $y_1$ and $y_2$ be positive random variables that denote the times of an event and a censoring occurrence, respectively. We assume that $y_1$ and $y_2$ are conditionally independent given the covariate vector $\bm{z}$. The time to finish an observation, i.e., the time to an event or censoring, can be expressed as $t=\min(y_1, y_2)$. From $y_1$ and $y_2$, we define $\delta$ as a random variable taking a value of $1$ for $y_1\le y_2$ (event) and $0$ for $y_1>y_2$ (censoring). For an experiment with $n$ subjects, let $\bm{t}\equiv(t_1,t_2,\ldots,t_n)^\T$ and $D([k^{(j-1)},k^{(j)}))\equiv\{i\mid \delta_i=1,\ t_i\in[k^{(j-1)},k^{(j)});\ i=1, 2, \ldots, n\}$ denote the time to an event or censoring and the set of subjects for which the event occurs in the period $[k^{(j-1)},k^{(j)})$, respectively. Then, the partial likelihood function is given by
\begin{align*}
L(\bm{\beta},\bm{k}; \bm{t})\equiv\prod_{j=1}^{m+1}
\prod_{i\in D([k^{(j-1)},k^{(j)}))}\frac{\exp(\bm{\beta}^{(j)\T}\bm{z}_i)}
			{\sum_{i'\in R(t_i)}\exp(\bm{\beta}^{(j)\T}\bm{z}_{i'})},
\end{align*}
where $R(t)$ is the risk set $\{i'\mid t<t_{i'}\}$ at time $t$. Also, we can express the log-partial likelihood function as
\begin{align*}
l(\bm{\beta},\bm{k}; \bm{t})\equiv\sum_{j=1}^{m+1}
\sum_{i\in D([k^{(j-1)},k^{(j)}))}
			\Bigg[\bm{\beta}^{(j)\T}\bm{z}_i-\log\Bigg\{\sum_{i'\in R(t_i)}\exp(\bm{\beta}^{(j)\T}\bm{z}_{i'})\Bigg\}\Bigg].
\end{align*}
The approach in \cite{Cox72} estimates the parameters by maximizing this log-partial likelihood function. However, we generalize that approach here to estimate the regression and change-point parameters by maximizing the regularized log-partial likelihood function with a ridge-type regularization term (\citealt{hoerl1970ridge}) for the model given by \eqref{CoxRegCP}. Specifically, we define the regularized log-partial likelihood function as
\begin{align}
l_\xi(\bm{\beta},\bm{k}; \bm{t})\equiv\sum_{j=1}^{m+1}
\sum_{i\in D([k^{(j-1)},k^{(j)}))}
			\Bigg[\bm{\beta}^{(j)\T}\bm{z}_i-\log\Bigg\{\sum_{i'\in R(t_i)}\exp(\bm{\beta}^{(j)\T}\bm{z}_{i'})\Bigg\}
			-\frac{\xi}{2}\bm{\beta}^{(j)\T}\bm{\beta}^{(j)}\Bigg],
\label{rpll}
\end{align}
and we estimate the parameters by maximizing this function, where $\xi$ is a regularization parameter. Hereafter, we denote $D^{(j)}\equiv D([k^{(j-1)},k^{(j)}))$ to simplify the notation. Let $\hat{\bm{\beta}}_{\bm{k}}$ be the $\bm{\beta}$ that maximizes the regularized log-partial likelihood function with the fixed change-points $\bm{k}$, i.e., $\hat{\bm{\beta}}_{\bm{k}}\equiv\argsup_{\bm{\beta}}l_\xi(\bm{\beta},\bm{k}; \bm{t})$. Also, we denote $\hat{\bm{k}}\equiv\argsup_{\bm{k}}l_\xi(\hat{\bm{\beta}}_{\bm{k}},\bm{k}; \bm{t})$ as the estimator of the change-point parameter. Then, the regression parameter estimator is given by $\hat{\bm{\beta}}=\hat{\bm{\beta}}_{\hat{\bm{k}}}$.

Next, we define $\bm{h}(t_i,\bm{\beta}^{(j)})$ and $\bm{H}(t_i,\bm{\beta}^{(j)})$ as a $p$-dimensional vector $\sum_{i'\in R(t_i)}\bm{z}_{i'}\exp(\bm{\beta}^{(j)\T}\bm{z}_{i'})/\allowbreak\sum_{i'\in R(t_i)}\exp(\bm{\beta}^{(j)\T}\bm{z}_{i'})$ and a $p\times p$ matrix $\sum_{i'\in R(t_i)}\bm{z}_{i'}\bm{z}_{i'}^\T\exp(\bm{\beta}^{(j)\T}\bm{z}_{i'})/\sum_{i'\in R(t_i)}\exp(\bm{\beta}^{(j)\T}\bm{z}_{i'})$, respectively. Then, for each $j\in\{1, 2, \ldots, m+1\}$, the first and second derivatives on $\bm{\beta}^{(j)}$ for the regularized log-partial likelihood function can be expressed as follows:
\begin{align*}
\frac{\partial}{\partial\bm{\beta}^{(j)}}l_\xi(\bm{\beta},\bm{k}; \bm{t})
= \sum_{i\in D^{(j)}}\{\bm{z}_i-\bm{h}(t_i,\bm{\beta}^{(j)})-\xi\bm{\beta}^{(j)}\},
\end{align*}
and
\begin{align*}
\frac{\partial^2}{\partial\bm{\beta}^{(j)}\partial\bm{\beta}^{(j)\T}}l_\xi(\bm{\beta},\bm{k}; \bm{t})
= -\sum_{i\in D^{(j)}}\{\bm{H}(t_i,\bm{\beta}^{(j)})-\bm{h}(t_i,\bm{\beta}^{(j)})\bm{h}(t_i,\bm{\beta}^{(j)})^\T-\xi\bm{I}_p\},
\end{align*}
where $\bm{I}_p$ is the $p$-dimensional identity matrix.

\subsection{Asymptotics for partial likelihood method}
\label{sec2_2}
For the case when there are no change-points, \cite{Tsiatis1981large} showed the consistency and asymptotic normality of an estimator based on the partial likelihood method for the regression parameter $\bm{\beta}$ in the Cox proportional hazards model, where the score function for the partial likelihood is expressed as a sum of independent random variables. In this subsection, we reveal the asymptotic behavior of the estimator $\hat{\bm{\beta}}_{\bm{k}}$ by maximizing the regularized log-partial likelihood based on \eqref{rpll}. Because the possible range of times to events or censoring occurrences $t_1,t_2,\ldots,t_n$ is $[0,T]$, the values of the change-points $k^{(1)},k^{(2)},\ldots,k^{(m)}$ are finite even when considering the asymptotic theories as $n$ increases. First, we define the following $p$-dimensional square matrices:
\begin{align*}
	&\bm{A}_\xi^{*(j)}(\bm{\beta}, \bm{k}) 
		\equiv \E\Bigg(\frac{1}{n}\Bigg[\sum_{i\in D^{(j)}}\{\bm{z}_i-\bm{h}(t_i,\bm{\beta}^{(j)})-\xi\bm{\beta}^{(j)}\}\Bigg]
			\Bigg[\sum_{i\in D^{(j)}}
			\{\bm{z}_i-\bm{h}(t_i,\bm{\beta}^{(j)})-\xi\bm{\beta}^{(j)}\}\Bigg]^\T\Bigg),
\end{align*}

\begin{align*}
	&\bm{B}_\xi^{*(j)}(\bm{\beta}, \bm{k})
		\equiv\E\Bigg[\frac{1}{n}\sum_{i\in D^{(j)}}\{\bm{H}(t_i,\bm{\beta}^{(j)})
			-\bm{h}(t_i,\bm{\beta}^{(j)})\bm{h}(t_i,\bm{\beta}^{(j)})^\T+\xi\bm{I}_p\}\Bigg].
\end{align*}
For simplicity, we denote $\bm{A}_\xi^{*(j)}\equiv \bm{A}_\xi^{*(j)}(\bm{\beta}_{\xi}^{*}, \bm{k}^*)$ and $\bm{B}_\xi^{*(j)}\equiv \bm{B}_\xi^{*(j)}(\bm{\beta}_{\xi}^{*}, \bm{k}^*)$, where $\bm{\beta}_{\xi}^{*}\equiv\argsup_{\bm{\beta}}\allowbreak\E\{l_\xi(\bm{\beta},\bm{k}^*; \bm{t})\}$. Then, from \cite{Tsiatis1981large}, we have 
\begin{align*}
	\bm{B}_\xi^{*(j)}=\bm{A}_\xi^{*(j)}+\xi^{*(j)}\bm{I}_{p},
\end{align*}
where $\xi^{*(j)}\equiv\E\{|D([k^{*(j-1)},k^{*(j)}))|\}\xi$. Note that $\E\{(\partial/\partial\bm{\beta}^{(j)})l_{\xi}(\bm{\beta}_{\xi}^{*},\bm{k}^*; \bm{t})\}=\E\allowbreak\{(\partial/\partial\bm{\beta}^{(j)})l(\bm{\beta}_{\xi}^{*},\allowbreak\bm{k}^*; \bm{t})\}-\xi^{*(j)}\bm{\beta}_{\xi}^{*(j)}=\bm{0}_p$, where $\bm{0}_{p}$ is a $p$-dimensional zero vector.

For a vector of finite values, $\bm{s}\equiv(s^{(1)}, s^{(2)}, \ldots, s^{(m)})^\T$, we set $k^{(j)}=k^{*(j)}+s^{(j)}/n$ for each $j\in\{1, 2, \ldots, m\}$. Then, similarly to \cite{Tsiatis1981large}, we can trivially show the consistency and asymptotic normality of $\hat{\bm{\beta}}_{\bm{k}}$, and we obtain
\begin{align}
\notag 
	&\hat{\bm{\beta}}^{(j)}_{\bm{k}}-\bm{\beta}_{\xi}^{*(j)}=\oP(1)
\end{align}
and
\begin{align}
	\label{AsyNorm}
	&\sqrt{n}(\hat{\bm{\beta}}^{(j)}_{\bm{k}}-\bm{\beta}_{\xi}^{*(j)})
		\stackrel{\rm d}{\to}
		{\rm N}\{\bm{0}_{p}, (\bm{A}_\xi^{*(j)}+\xi^{*(j)}\bm{I}_{p})^{-1}\bm{A}_\xi^{*(j)}(\bm{A}_\xi^{*(j)}+\xi^{*(j)}\bm{I}_{p})^{-1})\},
\end{align}
where ${\rm N}(\bm{\mu},\bm{\Sigma})$ is a multivariate normal distribution with mean vector $\bm{\mu}$ and variance-covariance matrix $\bm{\Sigma}$. In addition, the consistent estimators of $\bm{A}_\xi^{*(j)}$ and $\bm{B}_\xi^{*(j)}$ are obtained as follows:
\begin{align}
	\hat{\bm A}_\xi^{*(j)}(\hat{\bm{\beta}}, \hat{\bm{k}})		
	\equiv \frac{1}{n}\Bigg[\sum_{i\in\hat{D}^{(j)}}
			\{\bm{z}_i-\bm{h}(t_i,\hat{\bm{\beta}}^{(j)})-\xi\hat{\bm{\beta}}^{(j)}\}\Bigg]
			\Bigg[\sum_{i\in\hat{D}^{(j)}}
			\{\bm{z}_i-\bm{h}(t_i,\hat{\bm{\beta}}^{(j)})-\xi\hat{\bm{\beta}}^{(j)}\}\Bigg]^\T,
\label{defhatA}
\end{align}
\begin{align*}
	\hat{\bm B}_\xi^{*(j)}(\hat{\bm{\beta}}, \hat{\bm{k}})
		\equiv \frac{1}{n}\sum_{i\in\hat{D}^{(j)}}\{\bm{H}(t_i,\hat{\bm{\beta}}^{(j)})
			-\bm{h}(t_i,\hat{\bm{\beta}}^{(j)})\bm{h}(t_i,\hat{\bm{\beta}}^{(j)})^\T+\xi\bm{I}_p\},
\end{align*}
where $\hat{D}^{(j)}=D([\hat{k}^{(j-1)},\hat{k}^{(j)}))$.

\subsection{AIC for partial likelihood method}
\label{sec2_3}
In this subsection, we introduce the AIC for the Cox proportional hazards model when the partial likelihood is used, which was derived in \cite{xu2009using}. They used a risk function based on the Kullback-Leibler divergence between the true and estimated models, as in the case of conventional AIC-type information criteria. The asymptotic bias for each of the regression parameters has been shown to be 1. Later, we will derive AIC-type information criteria for the model given by \eqref{CoxRegCP} in the same way.

Let $(\hat{\bm{\beta}}_{\bm{t}},\hat{\bm{k}}_{\bm{t}})\equiv\argsup_{\bm{\beta},\bm{k}} l_\xi(\bm{\beta},\bm{k};\bm{t})$ be an estimator of $(\bm{\beta},\bm{k})$ based on survival time data $\bm{t}\equiv(t_1,t_2,\ldots,t_n)^\T$. In addition, by letting $\bm{u}=(u_1,u_2,\ldots,u_n)^{\T}$ be a copy of $\bm{t}$, i.e., letting $\bm{u}$ independently follow the same distribution as $\bm{t}$, we obtain a divergence $-2\E_{\bm{u}}\{l_\xi(\hat{\bm{\beta}}_{\bm{t}},\hat{\bm{k}}_{\bm{t}};\bm{u})\}$ based on the loss used in estimation, where $\E_{\bm{u}}$ denotes the expectation with respect to $\bm{u}$. Then, for an initial estimator, we take $-2$ times the maximum regularized log-partial likelihood $-2l_\xi(\hat{\bm{\beta}}_{\bm{t}},\hat{\bm{k}}_{\bm{t}};\bm{t})$, which can be bias-corrected by
\begin{align*}
	\E_{\bm{t}}[2l_\xi(\hat{\bm{\beta}}_{\bm{t}},\hat{\bm{k}}_{\bm{t}}; \bm{t})
		-\E_{\bm{u}}\{2l_\xi(\hat{\bm{\beta}}_{\bm{t}},\hat{\bm{k}}_{\bm{t}}; \bm{u})\}]
	= 2\E\{l_\xi(\hat{\bm{\beta}}_{\bm{t}},\hat{\bm{k}}_{\bm{t}}; \bm{t})
		-l_\xi(\hat{\bm{\beta}}_{\bm{u}},\hat{\bm{k}}_{\bm{u}}; \bm{t})\}.
\end{align*}
However, as this expectation cannot be given explicitly, we will evaluate the bias asymptotically, as with the conventional AIC. First, by defining $ll_\xi(\bm{\beta},\bm{k}; \bm{t})\equiv l_\xi(\bm{\beta},\bm{k}; \bm{t})-l_\xi(\bm{\beta}_{\xi}^{*},\bm{k}^*; \bm{t})$, $\hat{\bm{\beta}}_{\bm{k},\bm{u}}\equiv\argsup_{\bm{\beta}}ll_\xi(\bm{\beta},\bm{k}; \bm{u})$, and $\hat{ll}_\xi(\bm{k}; \bm{t},\bm{u})\equiv ll_\xi(\hat{\bm{\beta}}_{\bm{k},\bm{u}},\bm{k}; \bm{t})$, we express the bias as
\begin{align*}
	&2\E\{ll_\xi(\hat{\bm{\beta}}_{\bm{t}},\hat{\bm{k}}_{\bm{t}}; \bm{t})
		-ll_\xi(\hat{\bm{\beta}}_{\bm{u}},\hat{\bm{k}}_{\bm{u}}; \bm{t})\} \\
	&=2\E\bigg[\sup_{\bm{k}}ll_\xi(\hat{\bm{\beta}}_{\bm{k},\bm{t}},\bm{k}; \bm{t})
		-ll_\xi\bigg\{\hat{\bm{\beta}}
		_{\argsup_{\bm{k}}ll_\xi(\hat{\bm{\beta}}_{\bm{k},\bm{u}},\bm{k}; \bm{u}),\bm{u}},
		\argsup_{\bm{k}}ll_\xi(\hat{\bm{\beta}}_{\bm{k},\bm{u}},\bm{k}; \bm{u}); \bm{t}\bigg\}\bigg] \\
	&=2\E\bigg[\sup_{\bm{k}}\hat{ll}_\xi(\bm{k}; \bm{t},\bm{t})
		-\hat{ll}_\xi\bigg\{\argsup_{\bm{k}}\hat{ll}_\xi(\bm{k}; \bm{u},\bm{u}); \bm{t},\bm{u}\bigg\}\bigg].
\end{align*}
Also, by defining $b_\xi(\bm{k}^*,\bm{\beta}_{\xi}^{*})$ as the weak limit of $\sup_{\bm{k}\in K}\hat{ll}_\xi(\bm{k}; \bm{t},\bm{t})-\hat{ll}_\xi\{\argsup_{\bm{k}\in K}\hat{ll}_\xi(\bm{k}; \bm{u},\bm{u});\allowbreak\bm{t},\bm{u}\}$, we regard $2\E\{b_\xi(\bm{k}^*,\bm{\beta}_{\xi}^{*})\}$ as the asymptotic bias. Here, $K$ denotes the set such that $\hat{ll}_\xi(\bm{k}; \bm{t},\bm{t})$ is $\OP(1)$ or positive; that is, it denotes the set for which there exists some positive constant $M$ such that $\P\{\hat{ll}_\xi(\bm{k}; \bm{t},\bm{t})\allowbreak>-M\}$ does not converge to $0$. Then, we can say that
\begin{align}
	-2l_\xi(\hat{\bm{\beta}}_{\bm{t}},\hat{\bm{k}}_{\bm{t}}; \bm{t})+2\E\{b_\xi(\bm{k}^*,\bm{\beta}_{\xi}^{*})\}
	\label{zenbias}
\end{align}
is the AIC for the Cox proportional hazards model with change-points when using the regularized partial likelihood method. If there are no change-points and the regularization parameter $\xi$ is $0$, then this is the same as the AIC given by \cite{xu2009using}, where $\E\{b_{\xi}(\bm{k}^*,\bm{\beta}^*)\}=\E\{b_{\xi}(\bm{\beta}^*)\}$ with $\xi=0$ becomes the number of parameters in $\bm{\beta}$.

\section{Main results}
\label{sec3}
In this section, under the setting of Section \ref{sec2_1}, we use the asymptotic property obtained in Section \ref{sec2_2} to develop a novel information criterion by reevaluating the asymptotic bias according to the original AIC derivation method, which was introduced in Section \ref{sec2_3}.

\subsection{Evaluation of asymptotic bias} 
\label{sec3_1}
Let us set $k^{(j)}=k^{*(j)}+s^{(j)}/n$ for each $j\in\{1, 2, \ldots, m\}$. First, we consider the case where $\bm{s}=(s^{(1)},s^{(2)},\ldots,s^{(m)})$ is a vector with finite values. By using the first-order Taylor expansion of $(\partial/\partial\bm{\beta}^{(j)})l_\xi(\hat{\bm{\beta}}_{\bm{k}}, \bm{k}; \bm{t})=\bm{0}_p$ around $\hat{\bm{\beta}}_{\bm{k}}^{(j)}=\bm{\beta}_{\xi}^{*(j)}$, we have
\begin{align*}
	\bm{0}_p&=\sum_{i\in D^{(j)}}\{\bm{z}_i-\bm{h}(t_i,\bm{\beta}_{\xi}^{*(j)})-\xi\bm{\beta}_{\xi}^{*(j)}\}\\
		&\hphantom{{}={}}-\sum_{i\in D^{(j)}}
			\{\bm{H}(t_i,\bm{\beta}_{\xi}^{*(j)})-\bm{h}(t_i,\bm{\beta}_{\xi}^{*(j)})\bm{h}(t_i,\bm{\beta}_{\xi}^{*(j)})^\T+\xi \bm{I}_p\}
			(\hat{\bm{\beta}}_{\bm{k}}^{(j)}-\bm{\beta}_{\xi}^{*(j)})\{1+\oP(1)\}.
\end{align*}
Then, it holds that
\begin{align*}
	\hat{\bm{\beta}}_{\bm{k}}^{(j)}-\bm{\beta}_{\xi}^{*(j)}
		= \ & \Bigg[\sum_{i\in D^{(j)}}
			\{\bm{H}(t_i,\bm{\beta}_{\xi}^{*(j)})-\bm{h}(t_i,\bm{\beta}_{\xi}^{*(j)})\bm{h}(t_i,\bm{\beta}_{\xi}^{*(j)})^\T+\xi \bm{I}_p\}\Bigg]^{-1}
		\notag \\
		&\Bigg[\sum_{i\in D^{(j)}}\{\bm{z}_i-\bm{h}(t_i,\bm{\beta}_{\xi}^{*(j)})-\xi\bm{\beta}_{\xi}^{*(j)}\}\Bigg]\{1+\oP(1)\}.
\end{align*}
Next, supposing that $k^{(j-1)}\le k^{*(j-1)}$ and $k^{(j)}\le k^{*(j)}$, we have
\begin{align}\label{KOrderInBeta}
	\hat{\bm{\beta}}_{\bm{k}}^{(j)}-\hat{\bm{\beta}}_{\bm{k}^*}^{(j)}
		&=\Bigg[\sum_{i\in D^{(j)}}
			\{\bm{H}(t_i,\bm{\beta}_{\xi}^{*(j)})-\bm{h}(t_i,\bm{\beta}_{\xi}^{*(j)})\bm{h}(t_i,\bm{\beta}_{\xi}^{*(j)})^\T+\xi \bm{I}_p\}\Bigg]^{-1}
			\notag \\
			&\hphantom{{}={}}
			\Bigg[\sum_{i\in D([k^{(j-1)},k^{*(j-1)}))\cup D([k^{*(j-1)},k^{(j)}))}
			\{\bm{z}_i-\bm{h}(t_i,\bm{\beta}_{\xi}^{*(j)})-\xi\bm{\beta}_{\xi}^{*(j)}\}\Bigg]\{1+\oP(1)\}
			\notag \\
			&\hphantom{{}={}}
			-\Bigg[\sum_{i\in D^{*(j)}}
			\{\bm{H}(t_i,\bm{\beta}_{\xi}^{*(j)})-\bm{h}(t_i,\bm{\beta}_{\xi}^{*(j)})\bm{h}(t_i,\bm{\beta}_{\xi}^{*(j)})^\T+\xi \bm{I}_p\}\Bigg]^{-1}
			\notag \\
			&\hphantom{{}={}-{}}
			\Bigg[\sum_{i\in D([k^{*(j-1)},k^{(j)}))\cup D([k^{(j)},k^{*(j)}))}
			\{\bm{z}_i-\bm{h}(t_i,\bm{\beta}_{\xi}^{*(j)})-\xi\bm{\beta}_{\xi}^{*(j)}\}\Bigg]\{1+\oP(1)\}
			\notag \\
		&=\Bigg[\sum_{i\in D^{*(j)}}
			\{\bm{H}(t_i,\bm{\beta}_{\xi}^{*(j)})-\bm{h}(t_i,\bm{\beta}_{\xi}^{*(j)})\bm{h}(t_i,\bm{\beta}_{\xi}^{*(j)})^\T+\xi \bm{I}_p\}\Bigg]^{-1}
			\OP(1)\{1+\oP(1)\} \notag \\
		&=\OP(1/n).
\end{align}
Here, $D^{*(j)}\equiv D([k^{*(j-1)},k^{*(j)}))$, and even if the relationships between $k^{(j-1)}$ and $k^{*(j-1)}$ and between $k^{(j)}$ and $k^{*(j)}$ are different, \eqref{KOrderInBeta} holds. Therefore, we obtain $\hat{\bm{\beta}}_{\bm{k}}-\hat{\bm{\beta}}_{\bm{k}^*}=\OP(1/n)$. From \cite{Tsiatis1981large}, we also have $\hat{\bm{\beta}}_{\bm{k}^*}-\bm{\beta}_{\xi}^{*}=\OP(1/\sqrt{n})$, which implies that \eqref{AsyNorm} holds.

Next, by using Taylor expansion around $\bm{\beta}_{\xi}^{*(j)}=\hat{\bm{\beta}}_{\bm{k}}^{(j)}$ for the regularized log-partial likelihood function and \eqref{KOrderInBeta}, we have
\begin{align}
	\label{llTaylorBetaHatKBetaStar}
	&l_\xi(\hat{\bm{\beta}}_{\bm{k}},\bm{k}; \bm{t})-l_\xi(\bm{\beta}_{\xi}^{*},\bm{k}; \bm{t}) \notag \\
		&=\sum_{j=1}^{m+1}\Bigg(-(\bm{\beta}_{\xi}^{*(j)}-\hat{\bm{\beta}}_{\bm{k}}^{(j)})^\T
			\Bigg[\sum_{i\in D^{(j)}}\{\bm{z}_i-\bm{h}(t_i,\hat{\bm{\beta}}_{\bm{k}}^{(j)})
			-\xi\hat{\bm{\beta}}_{\bm{k}}^{(j)}\}\Bigg]+
			\notag \\
			&\hphantom{{}={}\sum_{j=1}^{m+1}\Bigg[}\frac{1}{2}(\bm{\beta}_{\xi}^{*(j)}-\hat{\bm{\beta}}_{\bm{k}}^{(j)})^\T\Bigg[\sum_{i\in D^{(j)}}
			\{\bm{H}(t_i,\hat{\bm{\beta}}_{\bm{k}}^{(j)})-\bm{h}(t_i,\hat{\bm{\beta}}_{\bm{k}}^{(j)})\bm{h}
			(t_i,\hat{\bm{\beta}}_{\bm{k}}^{(j)})^\T
			+\xi\bm{I}_p\}\Bigg](\bm{\beta}_{\xi}^{*(j)}-\hat{\bm{\beta}}_{\bm{k}}^{(j)})\Bigg) \notag \\
		&\hphantom{{}={}} +\oP(1) \notag \\
		&=\sum_{j=1}^{m+1}\Bigg(\frac{1}{2}(\bm{\beta}_{\xi}^{*(j)}-\hat{\bm{\beta}}_{\bm{k}^*}^{(j)})^\T
			\Bigg[\sum_{i\in D^{(j)}}\{\bm{H}(t_i,\hat{\bm{\beta}}_{\bm{k}^*}^{(j)})
				-\bm{h}(t_i,\hat{\bm{\beta}}_{\bm{k}^*}^{(j)})\bm{h}(t_i,\hat{\bm{\beta}}_{\bm{k}^*}^{(j)})^\T+\xi\bm{I}_p\}\Bigg] (\bm{\beta}_{\xi}^{*(j)}-\hat{\bm{\beta}}_{\bm{k}^*}^{(j)})\Bigg) \notag \\
		&\hphantom{{}={}} +\oP(1) \notag \\
		&=l_\xi(\hat{\bm{\beta}}_{\bm{k}^*},\bm{k}^*; \bm{t})-l_\xi(\bm{\beta}_{\xi}^{*},\bm{k}^*; \bm{t})+\oP(1).
\end{align}
Here, by defining
\begin{align*}
	Q_{\xi; k, \bm{t}}^{(j)} \equiv \ & I_{\{k<k^{*(j)}\}}\Bigg(\sum_{i\in D([k,k^{*(j)}))}\Bigg[(\bm{\beta}_{\xi}^{*(j+1)}-\bm{\beta}_{\xi}^{*(j)})^\T\bm{z}_i
			-\log\Bigg\{\frac{\sum_{i'\in R(t_i)}\exp(\bm{\beta}_{\xi}^{*(j+1)\T} \bm{z}_{i'})}
			{\sum_{i'\in R(t_i)}\exp(\bm{\beta}_{\xi}^{*(j)\T} \bm{z}_{i'})}\Bigg\}\notag \\
			& \phantom{I_{\{k<k^{*(j)}\}}\Bigg(}
			-\frac{\xi}{2}(\bm{\beta}_{\xi}^{*(j+1)\T}\bm{\beta}_{\xi}^{*(j+1)}-\bm{\beta}_{\xi}^{*(j)\T}\bm{\beta}_{\xi}^{*(j)})
\Bigg]\Bigg)\\
		& +I_{\{k>k^{*(j)}\}}\Bigg(\sum_{i\in D([k^{*(j)},k))}\Bigg[(\bm{\beta}_{\xi}^{*(j)}-\bm{\beta}_{\xi}^{*(j+1)})^\T\bm{z}_i
			-\log\Bigg\{\frac{\sum_{i'\in R(t_i)}\exp(\bm{\beta}_{\xi}^{*(j)\T} \bm{z}_{i'})}
			{\sum_{i'\in R(t_i)}\exp(\bm{\beta}_{\xi}^{*(j+1)\T} \bm{z}_{i'})}\Bigg\}\notag \\
			& \ \phantom{+I_{\{k>k^{*(j)}\}}\Bigg(}
			-\frac{\xi}{2}(\bm{\beta}_{\xi}^{*(j)\T}\bm{\beta}_{\xi}^{*(j)}-\bm{\beta}_{\xi}^{*(j+1)\T}\bm{\beta}_{\xi}^{*(j+1)})\Bigg]\Bigg)
\end{align*}
as a two-sided random walk with negative drift, we obtain
\begin{align}\label{llTaylorBetaHatKBetaHatKStar}
	&l_\xi(\hat{\bm{\beta}}_{\bm{k}},\bm{k}; \bm{t})-l_\xi(\hat{\bm{\beta}}_{\bm{k}^*},\bm{k}^*; \bm{t}) \notag \\
		&=l_\xi(\bm{\beta}_{\xi}^{*},\bm{k}; \bm{t})-l_\xi(\bm{\beta}_{\xi}^{*},\bm{k}^*; \bm{t})+\oP(1)
		=\sum_{j=1}^m Q_{\xi; k^{*(j)}+s^{(j)}/n, \bm{t}}^{(j)}+\oP(1)=\OP(1).
\end{align}
Furthermore, by using Taylor expansion around $\hat{\bm{\beta}}_{\bm{k}^*}=\bm{\beta}_{\xi}^{*}$ for the regularized log-partial likelihood, and from \eqref{AsyNorm} and \cite{Murphy2000profile}, the following holds:
\begin{align}\label{llTaylorBetaHatKStarBetaStarKStar}
	l_\xi(\hat{\bm{\beta}}_{\bm{k}^*},\bm{k}^*; \bm{t})-l_\xi(\bm{\beta}_{\xi}^{*},\bm{k}^*; \bm{t})
		&=\frac{1}{2}\sum_{j=1}^{m+1}\bm{\nu}_\xi^{(j)\T}\bm{\nu}_\xi^{*(j)}+\oP(1)=\OP(1),
\end{align}
where $\bm{\nu}_\xi^{*(j)}$ is a random variable vector distributed according to a multivariate normal distribution ${\rm N}\{\bm{0}_{p}, \bm{A}_\xi^{*(j)-1}(\bm{A}_\xi^{*(j)}+\xi^{*(j)}\bm{I}_{p})\}$. From \eqref{llTaylorBetaHatKBetaHatKStar} and \eqref{llTaylorBetaHatKStarBetaStarKStar}, we thus have
\begin{align*}
	\hat{ll}_\xi(\bm{k}; \bm{t},\bm{t})=\OP(1).
\end{align*}

Second, we consider the case where $\bm{s}$ is not a vector with finite values. In this case, there exists some $j$ such that $|k^{(j)}-k^{*(j)}|\to\infty$ ($n\to\infty$). Because it would be unnecessarily complicated to deal with this case in general, let us consider the following case for an index $j'$: 
\begin{align*}
	\bigg\{
	\begin{array}{l}
		k^{(j')}=k^{*(j')}+s^{(j')}/n,	\qquad 0>s^{(j')}\neq\O(1), \\
		k^{(j)}=k^{*(j)}+s^{(j)}/n, \qquad s^{(j)}=\O(1) \qquad (j\neq j').
	\end{array}
\end{align*}
In this case, $l_\xi(\hat{\bm{\beta}}_{\bm{k}},\bm{k}; \bm{t})-l_\xi(\hat{\bm{\beta}}_{\bm{k}^*},\bm{k}^*; \bm{t})$ can be decomposed into
\begin{align}
		&\sum_{i\in D([k^{*(j'-1)},k^{(j')}))}\Bigg[(\hat{\bm{\beta}}_{\bm{k}}^{(j')}-\hat{\bm{\beta}}_{\bm{k}^*}^{(j')})^{\T}\bm{z}_i
			-\log\Bigg\{\frac{\sum_{i'\in R(t_i)}\exp(\hat{\bm{\beta}}_{\bm{k}}^{(j')\T}\bm{z}_{i'})}
				{\sum_{i'\in R(t_i)}\exp(\hat{\bm{\beta}}_{\bm{k}^*}^{(j')\T}\bm{z}_{i'})}\Bigg\}
\notag\\
&\phantom{\sum_{i\in D([k^{*(j'-1)},k^{(j')}))}\Bigg[}
-\frac{\xi}{2}(\hat{\bm{\beta}}_{\bm{k}}^{(j')\T}\hat{\bm{\beta}}_{\bm{k}}^{(j')}
			-\hat{\bm{\beta}}_{\bm{k}^*}^{(j')\T}\hat{\bm{\beta}}_{\bm{k}^*}^{(j')})\Bigg]
\notag\\
		&+\sum_{i\in D([k^{(j')},k^{*(j')}))}\Bigg[(\hat{\bm{\beta}}_{\bm{k}}^{(j'+1)}-\hat{\bm{\beta}}_{\bm{k}^*}^{(j')})^{\T}\bm{z}_i
			-\log\Bigg\{\frac{\sum_{i'\in R(t_i)}\exp(\hat{\bm{\beta}}_{\bm{k}}^{(j'+1)\T}\bm{z}_{i'})}
				{\sum_{i'\in R(t_i)}\exp(\hat{\bm{\beta}}_{\bm{k}^*}^{(j')\T}\bm{z}_{i'})}\Bigg\}
\notag\\
&\phantom{+\sum_{i\in D([k^{(j')},k^{*(j')}))}\Bigg[}
-\frac{\xi}{2}(\hat{\bm{\beta}}_{\bm{k}}^{(j')\T}\hat{\bm{\beta}}_{\bm{k}}^{(j')}
			-\hat{\bm{\beta}}_{\bm{k}^*}^{(j')\T}\hat{\bm{\beta}}_{\bm{k}^*}^{(j')})\Bigg]
\notag\\
		&+\sum_{i\in D([k^{*(j')},k^{*(j'+1)}))}\Bigg[(\hat{\bm{\beta}}_{\bm{k}}^{(j'+1)}-\hat{\bm{\beta}}_{\bm{k}^*}^{(j'+1)})^{\T}\bm{z}_i
			-\log\Bigg\{\frac{\sum_{i'\in R(t_i)}\exp(\hat{\bm{\beta}}_{\bm{k}}^{(j'+1)\T}\bm{z}_{i'})}
				{\sum_{i'\in R(t_i)}\exp(\hat{\bm{\beta}}_{\bm{k}^*}^{(j'+1)\T}\bm{z}_{i'})}\Bigg\}
\notag\\
&\phantom{+\sum_{i\in D([k^{*(j')},k^{*(j'+1)}))}\Bigg[}
-\frac{\xi}{2}(\hat{\bm{\beta}}_{\bm{k}}^{(j'+1)\T}\hat{\bm{\beta}}_{\bm{k}}^{(j'+1)}
			-\hat{\bm{\beta}}_{\bm{k}^*}^{(j'+1)\T}\hat{\bm{\beta}}_{\bm{k}^*}^{(j'+1)})\Bigg]
\label{llRatioBetaHatKBetaHatKStar}
\end{align}
plus $\OP(1)$. By using Taylor expansion, the first sum reduces to
\begin{align}
(\hat{\bm{\beta}}_{\bm{k}}^{(j')}-\hat{\bm{\beta}}_{\bm{k}^*}^{(j')})^{\T}
\sum_{i\in D([k^{*(j'-1)},k^{(j')}))}
\Bigg[\bm{z}_i
			-\frac{\partial}{\partial\bm{\beta}^{(j')}}\log\Bigg\{\sum_{i'\in R(t_i)}\exp(\hat{\bm{\beta}}_{\bm{k}^*}^{(j')\T}\bm{z}_{i'})\Bigg\} 
-\xi\hat{\bm{\beta}}_{\bm{k}^*}^{(j')}\Bigg],
\notag
\end{align}
and the third sum reduces to
\begin{align}
(\hat{\bm{\beta}}_{\bm{k}}^{(j'+1)}-\hat{\bm{\beta}}_{\bm{k}^*}^{(j'+1)})^{\T}
		\sum_{i\in D([k^{*(j')},k^{*(j'+1)}))}\Bigg[\bm{z}_i
			-\frac{\partial}{\partial\bm{\beta}^{(j'+1)}}\log\Bigg\{\sum_{i'\in R(t_i)}\exp(\hat{\bm{\beta}}_{\bm{k}^*}^{(j'+1)\T}\bm{z}_{i'})\Bigg\} 
-\xi\hat{\bm{\beta}}_{\bm{k}^*}^{(j'+1)}\Bigg].
\notag
\end{align}
Each of these expressions is $\oP(s^{(j')})$, because $\hat{\bm{\beta}}_{\bm{k}}^{(j')}-\hat{\bm{\beta}}_{\bm{k}^*}^{(j')}=\OP(s^{(j')}/n)$, $\hat{\bm{\beta}}_{\bm{k}}^{(j'+1)}-\hat{\bm{\beta}}_{\bm{k}^*}^{(j'+1)}=\OP(s^{(j')}/n)$, $\hat{\bm{\beta}}_{\bm{k}^*}^{(j')}=\bm{\beta}_{\xi}^{*(j')}+\oP(1)$, and $\hat{\bm{\beta}}_{\bm{k}^*}^{(j'+1)}=\bm{\beta}_{\xi}^{*(j'+1)}+\oP(1)$. Lastly, the second sum can be written as
\begin{align}
\sum_{i\in D([k^{(j')},k^{*(j')}))}\Bigg[ & (\hat{\bm{\beta}}_{\bm{k}}^{(j'+1)}-\bm{\beta}_{\xi}^{*(j')})^{\T}\bm{z}_i
			-\log\Bigg\{\frac{\sum_{i'\in R(t_i)}\exp(\hat{\bm{\beta}}_{\bm{k}}^{(j'+1)\T}\bm{z}_{i'})}
				{\sum_{i'\in R(t_i)}\exp(\bm{\beta}_{\xi}^{*(j')\T}\bm{z}_{i'})}\Bigg\}
\notag \\
& -\frac{\xi}{2}(\hat{\bm{\beta}}_{\bm{k}}^{(j'+1)\T}\hat{\bm{\beta}}_{\bm{k}}^{(j'+1)}-\bm{\beta}_{\xi}^{*(j')\T}\bm{\beta}_{\xi}^{*(j')})\Bigg] + \oP(s^{(j')}),
\notag
\end{align}
whose expectation is negative and $\OP(s^{(j')})$ by the definition of $\bm{\beta}_{\xi}^{*(j')}$. From the above expressions, for any $M>0$, we have $\P\{l_\xi(\hat{\bm{\beta}}_{\bm{k}},\bm{k};\bm{t})-l_\xi(\hat{\bm{\beta}}_{\bm{k}^*},\bm{k}^*;\bm{t})>-M\}\to0$; it then follows from \eqref{llTaylorBetaHatKStarBetaStarKStar} that $\P\{\hat{ll}_\xi(\bm{k}; \bm{t},\bm{t})>-M\}\to0$ for any $M>0$. Thus, we obtain
\begin{align}
\hat{\bm{k}}-\bm{k}^*=\OP(1/n),
\label{forsm1}
\end{align}
which is consistent with the result in \cite{pons2002estimation}.

From the above derivation, we obtain $K=\{\bm{k}\mid k^{(j)}\in K^{(j)},\ j\in\{1,2,\ldots,m\}\}$, where $K^{(j)}=\{k\mid k-k^{*(j)}=\O(1/n)\}$. Therefore, from \eqref{llTaylorBetaHatKBetaHatKStar} and \eqref{llTaylorBetaHatKStarBetaStarKStar}, we have
\begin{align}
	&\sup_{\bm{k}\in K}\hat{ll}_\xi(\bm{k}; \bm{t},\bm{t})
		=\sum_{j=1}^m \sup_{k\in K^{(j)}}Q_{\xi; k,\bm{t}}^{(j)}
			+\frac{1}{2}\sum_{j=1}^{m+1}\bm{\nu}_\xi^{(j)\T}\bm{\nu}_\xi^{*(j)}+\oP(1) 
	\label{SupllHat}
\end{align}
and
\begin{align}
	&\argsup_{\bm{k}\in K}\hat{ll}_\xi(\bm{k}; \bm{u},\bm{u})
		=\bigg(\argsup_{k\in K^{(1)}}Q_{\xi; k,\bm{u}}^{(1)}, \argsup_{k\in K^{(2)}}Q_{\xi; k,\bm{u}}^{(2)},\ldots,
		\argsup_{k\in K^{(m)}}Q_{\xi; k,\bm{u}}^{(m)}\bigg)^\T+\oP(1).
	\label{ArgsupllHat}
\end{align}
In addition, by defining $\tilde{k}_{\bm{u}}^{(j)}\equiv\argsup_{k\in K^{(j)}}Q_{\xi; k,\bm{u}}^{(j)}$ and $\tilde{\bm{k}}_{\bm{u}}\equiv(\tilde{k}_{\bm{u}}^{(1)},\tilde{k}_{\bm{u}}^{(2)},\ldots,\tilde{k}_{\bm{u}}^{(m)})^\T$, we have
\begin{align}
	\hat{\bm{\beta}}_{\tilde{\bm{k}}_{\bm{u}},\bm{u}}-\hat{\bm{\beta}}_{\bm{k}^*,\bm{u}}=\OP(1/n)
\notag 
\end{align}
and
\begin{align}
	\hat{\bm{\beta}}_{\tilde{\bm{k}}_{\bm{u}},\bm{u}}-\bm{\beta}_{\xi}^{*}=\OP(1/\sqrt{n}).
\notag 
\end{align}
From these asymptotic properties and \cite{Murphy2000profile}, we obtain
\begin{align}
	\label{llTaylorBetaStarKTildeBetaHatKTilde}
	&l_\xi(\bm{\beta}_{\xi}^{*},\tilde{\bm{k}}_{\bm{u}}; \bm{t})
		-l_\xi(\hat{\bm{\beta}}_{\tilde{\bm{k}}_{\bm{u}},\bm{u}},\tilde{\bm{k}}_{\bm{u}}; \bm{t}) \notag \\
	&=\sum_{j=1}^{m+1}\Bigg(-(\hat{\bm{\beta}}_{\tilde{\bm{k}}_{\bm{u}},\bm{u}}^{(j)}-\bm{\beta}_{\xi}^{*(j)})^\T
		\Bigg[\sum_{i\in \tilde{D}_{\bm{u}}^{(j)}}\{\bm{z}_i-\bm{h}(t_i,\bm{\beta}_{\xi}^{*(j)})
			-\xi\bm{\beta}_{\xi}^{*(j)}\}\Bigg] 
\notag \\
		& \ \hphantom{=}
			-\frac{1}{2}(\hat{\bm{\beta}}_{\tilde{\bm{k}}_{\bm{u}},\bm{u}}^{(j)}-\bm{\beta}_{\xi}^{*(j)})^\T 
			\E\Bigg[\sum_{i\in \tilde{D}_{\bm{u}}^{(j)}}
			\{\bm{H}(t_i,\bm{\beta}_{\xi}^{*(j)})-\bm{h}(t_i,\bm{\beta}_{\xi}^{*(j)})\bm{h}(t_i,\bm{\beta}_{\xi}^{*(j)})^\T+\xi \bm{I}_p\}\Bigg]
			(\hat{\bm{\beta}}_{\tilde{\bm{k}}_{\bm{u}},\bm{u}}^{(j)}-\bm{\beta}_{\xi}^{*(j)}) \Bigg)
\notag \\
	& \ \hphantom{=} +\oP(1) \notag \\
	&=\sum_{j=1}^{m+1}\Bigg(-(\hat{\bm{\beta}}_{\bm{k}^*,\bm{u}}^{(j)}-\bm{\beta}_{\xi}^{*(j)})^\T
		\Bigg[\sum_{i\in D^{*(j)}}\{\bm{z}_i-\bm{h}(t_i,\bm{\beta}_{\xi}^{*(j)})
			-\xi\bm{\beta}_{\xi}^{*(j)}\}\Bigg] 
\notag \\
			& \ \hphantom{=}
-\frac{1}{2}(\hat{\bm{\beta}}_{\bm{k}^*,\bm{u}}^{(j)}-\bm{\beta}_{\xi}^{*(j)})^\T
			\E\Bigg[\sum_{i\in D^{*(j)}}
			\{\bm{H}(t_i,\bm{\beta}_{\xi}^{*(j)})-\bm{h}(t_i,\bm{\beta}_{\xi}^{*(j)})\bm{h}(t_i,\bm{\beta}_{\xi}^{*(j)})^\T+\xi \bm{I}_p\}\Bigg]
			(\hat{\bm{\beta}}_{\bm{k}^*,\bm{u}}^{(j)}-\bm{\beta}_{\xi}^{*(j)}) \Bigg)
	\notag \\
	& \ \hphantom{=} +\oP(1) \notag \\
	&=-\sum_{j=1}^{m+1}\Bigg((\hat{\bm{\beta}}_{\bm{k}^*,\bm{u}}^{(j)}-\bm{\beta}_{\xi}^{*(j)})^\T
		\Bigg[\sum_{i\in D^{*(j)}}\{\bm{z}_i-\bm{h}(t_i,\bm{\beta}_{\xi}^{*(j)})-\xi\bm{\beta}_{\xi}^{*(j)}\}\Bigg]
		+\frac{1}{2}\bm{\nu}_\xi^{(j)\T}\bm{\nu}_\xi^{*(j)}\Bigg)+\oP(1),
\end{align}
where $\tilde{D}_{\bm{u}}^{(j)}=D([\tilde{k}_{\bm{u}}^{(j-1)},\tilde{k}_{\bm{u}}^{(j)}))$. We also have
\begin{align}
	\label{llTaylorBetaStarKStarBetaStarKTilde}
	&l_\xi(\bm{\beta}_{\xi}^{*},\bm{k}^*; \bm{t})-l_\xi(\bm{\beta}_{\xi}^{*},\tilde{\bm{k}}_{\bm{u}}; \bm{t}) \notag \\
		&=\sum_{j=1}^m\Bigg\{I_{\{\tilde{k}_{\bm{u}}^{(j)}<k^{*(j)}\}}
			\Bigg(\sum_{i\in D([\tilde{k}_{\bm{u}}^{(j)},k^{*(j)}))}\Bigg[(\bm{\beta}_{\xi}^{*(j)}-\bm{\beta}_{\xi}^{*(j+1)})^\T\bm{z}_i
			-\log\Bigg\{\frac{\sum_{i'\in R(t_i)}\exp(\bm{\beta}_{\xi}^{*(j)\T} \bm{z}_{i'})}
			{\sum_{i'\in R(t_i)}\exp(\bm{\beta}_{\xi}^{*(j+1)\T} \bm{z}_{i'})}\Bigg\} \notag \\
			& \ \phantom{=\sum_{j=1}^m\Bigg\{I_{\{\tilde{k}_{\bm{u}}^{(j)}<k^{*(j)}\}}\Bigg(}
			-\frac{\xi}{2}(\bm{\beta}_{\xi}^{*(j)\T}\bm{\beta}_{\xi}^{*(j)}-\bm{\beta}_{\xi}^{*(j+1)\T}\bm{\beta}_{\xi}^{*(j+1)}) \Bigg] \Bigg)
\notag \\
		& \ \phantom{=}
		+I_{\{\tilde{k}_{\bm{u}}^{(j)}>k^{*(j)}\}}
			\Bigg(\sum_{i\in D([k^{*(j)},\tilde{k}_{\bm{u}}^{(j)}))}\Bigg[(\bm{\beta}_{\xi}^{*(j+1)}-\bm{\beta}_{\xi}^{*(j)})^\T\bm{z}_i
			-\log\Bigg\{\frac{\sum_{i'\in R(t_i)}\exp(\bm{\beta}_{\xi}^{*(j+1)\T} \bm{z}_{i'})}
			{\sum_{i'\in R(t_i)}\exp(\bm{\beta}_{\xi}^{*(j)\T} \bm{z}_{i'})}\Bigg\} \notag \\
			& \ \phantom{=+I_{\{\tilde{k}_{\bm{u}}^{(j)}>k^{*(j)}\}}\Bigg(}
			-\frac{\xi}{2}(\bm{\beta}_{\xi}^{*(j+1)\T}\bm{\beta}_{\xi}^{*(j+1)}-\bm{\beta}_{\xi}^{*(j)\T}\bm{\beta}_{\xi}^{*(j)})
			\Bigg]\Bigg)\Bigg\} \notag \\
	&=\sum_{j=1}^m Q_{\xi; \tilde{k}_{\bm{u}}^{(j)},\bm{t}}^{(j)}.
\end{align}
It thus follows that
\begin{align}
	\label{llHatKTilde}
	&\hat{ll}_\xi\bigg(\argsup_{\bm{k}\in K}\hat{ll}_\xi(\bm{k}; \bm{u},\bm{u}); \bm{t},\bm{u}\bigg) \notag \\
	&=\hat{ll}_\xi(\tilde{\bm{k}}_{\bm{u}}; \bm{t},\bm{u})+\oP(1) \notag \\
	&=-\sum_{j=1}^m Q_{\xi; \tilde{k}_{\bm{u}}^{(j)},\bm{t}}^{(j)} + \sum_{j=1}^{m+1}\Bigg((\hat{\bm{\beta}}_{\bm{k}^*,\bm{u}}^{(j)}-\bm{\beta}_{\xi}^{*(j)})^\T
		\Bigg[\sum_{i\in D^{*(j)}}\{\bm{z}_i-\bm{h}(t_i,\bm{\beta}_{\xi}^{*(j)})-\xi\bm{\beta}_{\xi}^{*(j)}\}\Bigg]
		-\frac{1}{2}\bm{\nu}_\xi^{(j)\T}\bm{\nu}_\xi^{*(j)}\Bigg)
\notag \\
	& \ \phantom{=} + \oP(1).
\end{align}
Finally, from \eqref{SupllHat} and \eqref{llHatKTilde}, we can obtain the following theorem.

\begin{theorem}
Under condition \eqref{ConditionBeta}, the asymptotic bias in \eqref{zenbias} is given by
	\begin{align}
		&\E\{b_\xi(\bm{k}^*,\bm{\beta}_{\xi}^{*})\}
		\notag \\
		&=\sum_{j=1}^m\E\bigg(\sup_{k\in K^{(j)}}Q_{\xi; k,\bm{t}}^{(j)}
					+Q_{\xi; \argsup_{k\in K^{(j)}}Q_{\xi; k,\bm{u}}^{(j)},\bm{t}}^{(j)}\bigg)
		+\sum_{j=1}^{m+1}\text{tr}\{\bm{A}_\xi^{*(j)}(\bm{A}_\xi^{*(j)}+\xi^{*(j)} \bm{I}_{p})^{-1}\}.
		\label{tt1}
	\end{align}
	\label{theorem1}
\end{theorem}

\noindent
We can regard the first and second terms on the right side of \eqref{tt1} as the biases for the change-point parameters $\bm{k}$ and the regression parameters $\bm{\beta}$, respectively.

\subsection{Explicit expression of asymptotic bias} 
\label{sec3_2}
In the AIC for regular statistical models, the penalty is 2 for each parameter, regardless of whether its true value is a constant or converges to any value. In other words, it does not matter which setting is considered. On the other hand, in the AIC for conventional change-point models, the penalty depends on the setting. In particular, the evaluation of the first term on the right side of \eqref{tt1} depends on whether $\bm{\beta}^{*(j+1)}-\bm{\beta}^{*(j)}$ is a constant vector or converges to $\bm{0}_p$.

Here, we deem the latter case more important and natural. If $\bm{\beta}^{*(j+1)}-\bm{\beta}^{*(j)}$ is a constant vector even in asymptotics, a clear change is expected to exist. In such a setting, the first term of the information criterion, i.e., the goodness-of-fit term, almost entirely determines the model selection result, and the bias evaluation of the second term is less important. For cases in which it cannot be determined at first sight whether there are change-points, we need a more accurate evaluation of the second term, and the assumption of $\bm{\beta}^{*(j+1)}-\bm{\beta}^{*(j)}\to\bm{0}_p$ reflects such a case. Even if the existence of change-points is suspected at first glance, their existence is not absolute as long as the data size is finite. The assumption that $\bm{\beta}^{*(j+1)}-\bm{\beta}^{*(j)}$ is a constant vector leads to asymptotic approximations that are too biased toward the existence of changes. Therefore, we consider it more natural to assume that $\bm{\beta}^{*(j+1)}-\bm{\beta}^{*(j)}$ converges to $\bm{0}_p$.

From the above discussion, as in Section 1.5 of \cite{CsoHor97}, when estimating the parameters by maximizing the regularized log-partial likelihood function, we assume the following condition:
\begin{align}
	\bm{\beta}_{\xi}^{*(j+1)}-\bm{\beta}_{\xi}^{*(j)}=\bm{\bm{\Delta}}_{\bm{\beta}_{\xi}^{*}}^{(j)}/\sqrt{\alpha_n} 
	\qquad (j\in\{1,2,\ldots,m\}), \qquad \O(1)\neq\alpha_n=\o(n),
	\label{condition}
\end{align}
where $\bm{\bm{\Delta}}_{\bm{\beta}_{\xi}^{*}}^{(j)}$ is a constant vector. Hence, we examine the asymptotic behavior of the change-point estimator under condition \eqref{condition}. First, similarly to the derivation of \eqref{forsm1}, in the case where $\bm{k}=\bm{k}^*+\alpha_n\bm{s}/n$ and $\bm{s}$ is a vector with finite values, we obtain $\hat{\bm{\beta}}_{\bm{k}}-\hat{\bm{\beta}}_{\bm{k}^*}=\OP(\sqrt{\alpha_n}/n)$ instead of \eqref{KOrderInBeta}. Then, from \eqref{llTaylorBetaHatKBetaStar}, \eqref{llTaylorBetaHatKBetaHatKStar}, and \eqref{llTaylorBetaHatKStarBetaStarKStar}, $\hat{ll}_\xi(\bm{k}; \bm{t},\bm{t})=\OP(1)$ holds. On the other hand, for the case where $\bm{s}$ is not a vector with finite values, as in the derivation of \eqref{forsm1}, let us consider the case where $k^{(j')}=k^{*(j')}+\alpha_ns^{(j')}/n,\ 0>s^{(j')}\neq\O(1)$, and $k^{(j)}=k^{*(j)}+\alpha_ns^{(j)}/n,\ s^{(j)}=\O(1)$, for $j\neq j'$. In this case, again via \eqref{llRatioBetaHatKBetaHatKStar}, we have $\P\{l_\xi(\hat{\bm{\beta}}_{\bm{k}},\bm{k}; \bm{t})-l_\xi(\hat{\bm{\beta}}_{\bm{k}^*},\bm{k}^*; \bm{t})>-M\}\to0$ for any $M>0$. Then, we can show that $\P\{\hat{ll}_\xi(\bm{k}; \bm{t},\bm{t})>-M\}\to 0$ by combining it with \eqref{llTaylorBetaHatKStarBetaStarKStar}. Finally, from these derivations, we obtain $\hat{\bm{k}}-\bm{k}^*=\OP(\alpha_n/n)$ instead of \eqref{forsm1}, and it can be seen that $K=\{\bm{k}\mid k^{(j)}\in K^{(j)},\ j\in\{1,2,\ldots,m\}\}$, where $K^{(j)}=\{k\mid k-k^{*(j)}=\O(\alpha_n/n)\}$, thus yielding \eqref{SupllHat}, \eqref{ArgsupllHat}, \eqref{llTaylorBetaStarKTildeBetaHatKTilde}, \eqref{llTaylorBetaStarKStarBetaStarKTilde}, and \eqref{llHatKTilde}.

Hereafter, letting $\bm{s}$ be a vector with finite values, we assume that $\bm{k}=\bm{k}^*+\alpha_n\bm{s}/n$. Under condition \eqref{condition}, $Q_{\xi; k^{*(j)}+\alpha_ns^{(j)}/n, \bm{t}}^{(j)}$ can be written as
\begin{align}
	\label{QKStarAlpha}
	& I_{\{s^{(j)}<0\}}\Bigg(\frac{1}{\sqrt{\alpha_n}} \bm{\bm{\Delta}}_{\bm{\beta}_{\xi}^{*}}^{(j)\T}
			\Bigg[\sum_{i\in D_{1\alpha_n}^{*(j)}}
			\{\bm{z}_i-\bm{h}(t_i,\bm{\beta}_{\xi}^{*(j)})-\xi\bm{\beta}_{\xi}^{*(j)}\} \notag \\
	& \phantom{I_{\{s^{(j)}<0\}}\Bigg(}
	-\frac{1}{2\alpha_n} \bm{\bm{\Delta}}_{\bm{\beta}_{\xi}^{*}}^{(j)\T}
			\Bigg[\sum_{i\in D_{1\alpha_n}^{*(j)}}
			\{\bm{H}(t_i,\bm{\beta}_{\xi}^{*(j)})-\bm{h}(t_i,\bm{\beta}_{\xi}^{*(j)})\bm{h}(t_i,\bm{\beta}_{\xi}^{*(j)})^\T
			+\xi\bm{I}_p\}\Bigg]\bm{\bm{\Delta}}_{\bm{\beta}_{\xi}^{*}}^{(j)}\Bigg) \notag \\
	& +I_{\{s^{(j)}>0\}}\Bigg(\frac{1}{\sqrt{\alpha_n}}\bm{\bm{\Delta}}_{\bm{\beta}_{\xi}^{*}}^{(j)\T}
			\Bigg[\sum_{i\in D_{2\alpha_n}^{*(j)}}
			\{\bm{z}_i-\bm{h}(t_i,\bm{\beta}_{\xi}^{*(j+1)})-\xi\bm{\beta}_{\xi}^{*(j+1)}\}\Bigg] \notag \\
	& \ \phantom{+I_{\{s^{(j)}>0\}}\Bigg(}
	-\frac{1}{2\alpha_n} \bm{\bm{\Delta}}_{\bm{\beta}_{\xi}^{*}}^{(j)\T}
			\Bigg[\sum_{i\in D_{2\alpha_n}^{*(j)}}
			\{\bm{H}(t_i,\bm{\beta}_{\xi}^{*(j+1)})-\bm{h}(t_i,\bm{\beta}_{\xi}^{*(j+1)})\bm{h}(t_i,\bm{\beta}_{\xi}^{*(j+1)})^\T
			+\xi\bm{I}_p\}\Bigg]\bm{\bm{\Delta}}_{\bm{\beta}_{\xi}^{*}}^{(j)}\Bigg)
\end{align}
plus $\oP(1)$, where $D_{1\alpha_n}^{*(j)}\equiv D([k^{*(j)}+\alpha_ns^{(j)}/n,k^{*(j)}))$ and $D_{2\alpha_n}^{*(j)}\equiv D([k^{*(j)},k^{*(j)}+\alpha_ns^{(j)}/\allowbreak n))$. Let $\{W_s\}_{s\in\mathbb{R}}$ denote two-sided standard Brownian motion with $\E(W_s)=0$ and $\V(W_s)=|s|$; then, we obtain
\begin{align}
	\label{Sigma1}
	&\frac{1}{\sqrt{\alpha_n}} \bm{\Delta}_{\bm{\beta}_{\xi}^{*}}^{(j)\T}
			\Bigg[\sum_{i\in D_{1\alpha_n}^{*(j)}}
			\{\bm{z}_i-\bm{h}(t_i,\bm{\beta}_{\xi}^{*(j)})-\xi\bm{\beta}_{\xi}^{*(j)}\}\Bigg]
\stackrel{\rm d}{\to}
		(\bm{\Delta}_{\bm{\beta}_{\xi}^{*}}^{(j)\T}\bm{A}_\xi^{*(j)}\bm{\Delta}_{\bm{\beta}_{\xi}^{*}}^{(j)})^{1/2}W_s,
\end{align}
\begin{align}
	\label{C1}
	&\frac{1}{2\alpha_n} \bm{\Delta}_{\bm{\beta}_{\xi}^{*}}^{(j)\T}
		\Bigg[\sum_{i\in D_{1\alpha_n}^{*(j)}}
		\{\bm{H}(t_i,\bm{\beta}_{\xi}^{*(j)})-\bm{h}(t_i,\bm{\beta}_{\xi}^{*(j)})\bm{h}(t_i,\bm{\beta}_{\xi}^{*(j)})^\T
		+\xi\bm{I}_p\}\Bigg]\bm{\Delta}_{\bm{\beta}_{\xi}^{*}}^{(j)} \notag \\
		&\stackrel{\rm p}{\to}
		\frac{1}{2}\bm{\Delta}_{\bm{\beta}_{\xi}^{*}}^{(j)\T}(\bm{A}_\xi^{*(j)}+\xi^{*(j)}\bm{I}_p)\bm{\Delta}_{\bm{\beta}_{\xi}^{*}}^{(j)}|s|,
\end{align}
\begin{align}
	\label{Sigma2}
	&\frac{1}{\sqrt{\alpha_n}} \bm{\Delta}_{\bm{\beta}_{\xi}^{*}}^{(j)\T}
		\Bigg[\sum_{i\in D_{2\alpha_n}^{*(j)}}
			\{\bm{z}_i-\bm{h}(t_i,\bm{\beta}_{\xi}^{*(j+1)})-\xi\bm{\beta}_{\xi}^{*(j+1)}\}\Bigg]
		\stackrel{\rm d}{\to}
		(\bm{\Delta}_{\bm{\beta}_{\xi}^{*}}^{(j)\T}\bm{A}_\xi^{*(j+1)}\bm{\Delta}_{\bm{\beta}_{\xi}^{*}}^{(j)})^{1/2}W_s,
\end{align}
and
\begin{align}
	\label{C2}
	&\frac{1}{2\alpha_n} \bm{\Delta}_{\bm{\beta}_{\xi}^{*}}^{(j)\T}
		\Bigg[\sum_{i\in D_{2\alpha_n}^{*(j)}}
		\{\bm{H}(t_i,\bm{\beta}_{\xi}^{*(j+1)})-\bm{h}(t_i,\bm{\beta}_{\xi}^{*(j+1)})\bm{h}(t_i,\bm{\beta}_{\xi}^{*(j+1)})^\T
			+\xi\bm{I}_p\}\Bigg]\bm{\Delta}_{\bm{\beta}_{\xi}^{*}}^{(j)} \notag \\
		&\stackrel{\rm p}{\to}
		\frac{1}{2}\bm{\Delta}_{\bm{\beta}_{\xi}^{*}}^{(j)\T}(\bm{A}_\xi^{*(j+1)}+\xi^{*(j+1)}\bm{I}_p)\bm{\Delta}_{\bm{\beta}_{\xi}^{*}}^{(j)}|s|.
\end{align}
Next, let $V_s(\tau_1,\tau_2,\sigma_1,\sigma_2)$ denote Brownian motion extending to both sides with drift coefficients of $\tau_1$ and $\tau_2$ and diffusion coefficients of $\sigma_1$ and $\sigma_2$; that is, we define $V_s(\tau_1,\tau_2,\sigma_1,\sigma_2)$ as $-\tau_1|s|+\sigma_1 W_s$ when $s<0$, and as $-\tau_2|s|+\sigma_2 W_s$ when $s\ge 0$. In addition, let
\begin{align*}
V_{\xi; s}^{*(j)} 
	\equiv V_s\bigg\{&\frac{1}{2}\bm{\Delta}_{\bm{\beta}_{\xi}^{*}}^{(j)\T}(\bm{A}_\xi^{*(j)}+\xi^{*(j)}\bm{I}_p)\bm{\Delta}_{\bm{\beta}_{\xi}^{*}}^{(j)},
							\frac{1}{2}\bm{\Delta}_{\bm{\beta}_{\xi}^{*}}^{(j)\T}(\bm{A}_\xi^{*(j+1)}+\xi^{*(j+1)}\bm{I}_p)\bm{\Delta}_{\bm{\beta}_{\xi}^{*}}^{(j)}, \notag \\
					&	(\bm{\Delta}_{\bm{\beta}_{\xi}^{*}}^{(j)\T}\bm{A}_\xi^{*(j)}\bm{\Delta}_{\bm{\beta}_{\xi}^{*}}^{(j)})^{1/2},
							(\bm{\Delta}_{\bm{\beta}_{\xi}^{*}}^{(j)\T}\bm{A}_\xi^{*(j+1)}\bm{\Delta}_{\bm{\beta}_{\xi}^{*}}^{(j)})^{1/2}\bigg\}.
\end{align*}
Then, from \eqref{QKStarAlpha}, \eqref{Sigma1}, \eqref{C1}, \eqref{Sigma2}, and \eqref{C2}, the following holds:
\begin{align*}
	Q_{\xi; k^*+\alpha_ns/n, \bm{t}}^{(j)}\stackrel{\rm d}{\to}V_{\xi; s}^{*(j)}.
\end{align*}
Thus, we have
\begin{align}
	\label{SupQ}
	&\sup_{k\in K^{(j)}}Q_{\xi; k,\bm{t}}^{(j)}
		\stackrel{\rm d}{\to}
		\sup_{s\in\mathbb{R}}V_{\xi; s}^{*(j)}
\end{align}
and
\begin{align}
	\label{Qargsup}
	&Q_{\xi; \argsup_{k\in K^{(j)}}Q_{\xi; k,\bm{u}}^{(j)},\bm{t}}^{(j)}
		\stackrel{\rm d}{\to}
		V_{\xi; \argsup_{s\in\mathbb{R}}V'^{*(j)}_{\xi; s}}^{*(j)}
\end{align}
as consequences, where $V'^{*(j)}_{\xi; s}$ is a copy of $V_{\xi; s}^{*(j)}$.

To evaluate these expectations, we use the results of \cite{BhaBro76} and \cite{She79}. First, by using the equality
\begin{align*}
	\P\bigg\{\sup_{s>0}(W_s-a_2s)>a_1\bigg\}=\exp(-2a_1a_2),
\end{align*}
which holds for positive constants $a_1$ and $a_2$, we obtain
\begin{align}
	&\E\bigg(\sup_{s\in\mathbb{R}}V_{\xi; s}^{*(j)}\bigg) \notag \\
	&=\int_0^\infty\P\bigg(\sup_{s\in\mathbb{R}}V_{\xi; s}^{*(j)}>a\bigg){\rm d}a \notag \\
	&=\int_0^\infty\Bigg[
		\exp\Bigg\{\frac{-\bm{\Delta}_{\bm{\beta}_{\xi}^{*}}^{(j)\T}(\bm{A}_\xi^{*(j)}+\xi^{*(j)}\bm{I}_p)\bm{\Delta}_{\bm{\beta}_{\xi}^{*}}^{(j)}}
			{\bm{\Delta}_{\bm{\beta}_{\xi}^{*}}^{(j)\T}\bm{A}_\xi^{*(j)}\bm{\Delta}_{\bm{\beta}_{\xi}^{*}}^{(j)}}a\Bigg\}
		+\exp\Bigg\{\frac{-\bm{\Delta}_{\bm{\beta}_{\xi}^{*}}^{(j)\T}(\bm{A}_\xi^{*(j+1)}+\xi^{*(j+1)}\bm{I}_p)\bm{\Delta}_{\bm{\beta}_{\xi}^{*}}^{(j)}}
			{\bm{\Delta}_{\bm{\beta}_{\xi}^{*}}^{(j)\T}\bm{A}_\xi^{*(j+1)}\bm{\Delta}_{\bm{\beta}_{\xi}^{*}}^{(j)}}a\Bigg\} \notag \\
		& \hphantom{=}
		-\exp\Bigg\{\frac{-\bm{\Delta}_{\bm{\beta}_{\xi}^{*}}^{(j)\T}(\bm{A}_\xi^{*(j)}+\xi^{*(j)}\bm{I}_p)\bm{\Delta}_{\bm{\beta}_{\xi}^{*}}^{(j)\T}
			\bm{A}_\xi^{*(j+1)}
			-\bm{\Delta}_{\bm{\beta}_{\xi}^{*}}^{(j)\T}(\bm{A}_\xi^{*(j+1)}+\xi^{*(j+1)}\bm{I}_p)\bm{\Delta}_{\bm{\beta}_{\xi}^{*}}^{(j)\T}\bm{A}_\xi^{*(j)}}
			{\bm{\Delta}_{\bm{\beta}_{\xi}^{*}}^{(j)\T}\bm{A}_\xi^{*(j)}\bm{\Delta}_{\bm{\beta}_{\xi}^{*}}^{(j)\T}\bm{A}_\xi^{*(j+1)}}
			a\Bigg\}\Bigg]\text{d}a 
\notag\\
	& = \bm{C}(\bm{A}_\xi^{*(j)}, \bm{A}_\xi^{*(j)}+\xi^{*(j)}\bm{I}_p),
	\label{BaBaS}
\end{align}
where
\begin{align*}
		\bm{C}(\bm{A}^{(j)\dagger}, \bm{A}^{(j)\ddagger})
			&=\{(\bm{\Delta}_{\bm{\beta}_{\xi}^{*}}^{(j)\T}\bm{A}^{(j)\ddagger}\bm{\Delta}_{\bm{\beta}_{\xi}^{*}}^{(j)}
				\bm{\Delta}_{\bm{\beta}_{\xi}^{*}}^{(j)\T}\bm{A}^{(j+1)\dagger}\bm{\Delta}_{\bm{\beta}_{\xi}^{*}}^{(j)})^{2} 
					+(\bm{\Delta}_{\bm{\beta}_{\xi}^{*}}^{(j)\T}\bm{A}^{(j+1)\ddagger}\bm{\Delta}_{\bm{\beta}_{\xi}^{*}}^{(j)}
					\bm{\Delta}_{\bm{\beta}_{\xi}^{*}}^{(j)\T}\bm{A}^{(j+1)\dagger}\bm{\Delta}_{\bm{\beta}_{\xi}^{*}}^{(j)})^{2}
\notag \\
				& \hphantom{=[\{}
					+\bm{\Delta}_{\bm{\beta}_{\xi}^{*}}^{(j)\T}\bm{A}^{(j)\ddagger}\bm{\Delta}_{\bm{\beta}_{\xi}^{*}}^{(j)}
					\bm{\Delta}_{\bm{\beta}_{\xi}^{*}}^{(j)\T}\bm{A}^{(j+1)\ddagger}\bm{\Delta}_{\bm{\beta}_{\xi}^{*}}^{(j)}
					\bm{\Delta}_{\bm{\beta}_{\xi}^{*}}^{(j)\T}\bm{A}^{(j)\dagger}\bm{\Delta}_{\bm{\beta}_{\xi}^{*}}^{(j)}
					\bm{\Delta}_{\bm{\beta}_{\xi}^{*}}^{(j)\T}\bm{A}^{(j+1)\dagger}\bm{\Delta}_{\bm{\beta}_{\xi}^{*}}^{(j)}\}
					\notag \\
				& \ \hphantom{=}/ \ \{\bm{\Delta}_{\bm{\beta}_{\xi}^{*}}^{(j)\T}\bm{A}^{(j)\ddagger}\bm{\Delta}_{\bm{\beta}_{\xi}^{*}}^{(j)}
				\bm{\Delta}_{\bm{\beta}_{\xi}^{*}}^{(j)\T}\bm{A}^{(j+1)\ddagger}\bm{\Delta}_{\bm{\beta}_{\xi}^{*}}^{(j)}
\notag \\
				&\hphantom{=/\ [}
				(\bm{\Delta}_{\bm{\beta}_{\xi}^{*}}^{(j)\T}\bm{A}^{(j)\ddagger}\bm{\Delta}_{\bm{\beta}_{\xi}^{*}}^{(j)}
				\bm{\Delta}_{\bm{\beta}_{\xi}^{*}}^{(j)\T}\bm{A}^{(j+1)\dagger}\bm{\Delta}_{\bm{\beta}_{\xi}^{*}}^{(j)} 
				+\bm{\Delta}_{\bm{\beta}_{\xi}^{*}}^{(j)\T}\bm{A}^{(j+1)\ddagger}\bm{\Delta}_{\bm{\beta}_{\xi}^{*}}^{(j)}
				\bm{\Delta}_{\bm{\beta}_{\xi}^{*}}^{(j)\T}\bm{A}^{(j)\dagger}\bm{\Delta}_{\bm{\beta}_{\xi}^{*}}^{(j)})\}.
\end{align*}
Next, we use the fact that the probability density function of $\argsup_{s\in\mathbb{R}}V_s(\tau_1,\tau_2,\sigma_1,\sigma_2)$ is given by $g(-s\mid \tau_1/\sigma_1,\allowbreak\tau_2\sigma_1/\sigma_2^2)$ when $s\le 0$ and by $g(s\mid \tau_2/\sigma_2,\tau_1\sigma_2/\sigma_1^2)$ when $s\ge 0$, where
\begin{align*}
	g(s\mid a_1,a_2)
		\equiv 2a_1(a_1+2a_2)\exp\{2a_2(a_1+a_2)s\}\Phi\{-(a_1+2a_2)\sqrt{s}\}-2a_1^2\Phi(-a_1\sqrt{s}).
\end{align*}
Then, we have
\begin{align}
	&\E\bigg(V_{\xi; \argsup_{s\in\mathbb{R}}V'^{*(j)}_{\xi; s}}^{*(j)}\bigg)
	\notag\\
	&=\int_0^\infty \frac{1}{2}sg\Bigg[s\;\Bigg|\;
		\frac{\bm{\Delta}_{\bm{\beta}_{\xi}^{*}}^{(j)\T}(\bm{A}_\xi^{*(j)}+\xi^{*(j)}\bm{I}_p)\bm{\Delta}_{\bm{\beta}_{\xi}^{*}}^{(j)}}
			{2(\bm{\Delta}_{\bm{\beta}_{\xi}^{*}}^{(j)\T}\bm{A}_\xi^{*(j)}\bm{\Delta}_{\bm{\beta}_{\xi}^{*}}^{(j)})^{1/2}},
	\notag \\
	& \ \hphantom{=\int_0^\infty}
		\frac{\bm{\Delta}_{\bm{\beta}_{\xi}^{*}}^{(j)\T}(\bm{A}_\xi^{*(j+1)}+\xi^{*(j+1)}\bm{I}_p)\bm{\Delta}_{\bm{\beta}_{\xi}^{*}}^{(j)}
			(\bm{\Delta}_{\bm{\beta}_{\xi}^{*}}^{(j)\T}\bm{A}_\xi^{*(j)}\bm{\Delta}_{\bm{\beta}_{\xi}^{*}}^{(j)})^{1/2}}
			{2\bm{\Delta}_{\bm{\beta}_{\xi}^{*}}^{(j)\T}\bm{A}_\xi^{*(j+1)}\bm{\Delta}_{\bm{\beta}_{\xi}^{*}}^{(j)}}\Bigg]
			\bm{\Delta}_{\bm{\beta}_{\xi}^{*}}^{(j)\T}(\bm{A}_\xi^{*(j)}+\xi^{*(j)}\bm{I}_p)\bm{\Delta}_{\bm{\beta}_{\xi}^{*}}^{(j)}\text{d}s \notag \\
		& \ \hphantom{=}
		+\int_0^\infty \frac{1}{2}sg\Bigg[s\;\Bigg|\;
			\frac{\bm{\Delta}_{\bm{\beta}_{\xi}^{*}}^{(j)\T}(\bm{A}_\xi^{*(j+1)}+\xi^{*(j+1)}\bm{I}_p)\bm{\Delta}_{\bm{\beta}_{\xi}^{*}}^{(j)}}
				{2(\bm{\Delta}_{\bm{\beta}_{\xi}^{*}}^{(j)\T}\bm{A}_\xi^{*(j+1)}\bm{\Delta}_{\bm{\beta}_{\xi}^{*}}^{(j)})^{1/2}},
	\notag \\
	& \ \hphantom{=+\int_0^\infty}
			\frac{\bm{\Delta}_{\bm{\beta}_{\xi}^{*}}^{(j)\T}(\bm{A}_\xi^{*(j)}+\xi^{*(j)}\bm{I}_p)\bm{\Delta}_{\bm{\beta}_{\xi}^{*}}^{(j)}
				(\bm{\Delta}_{\bm{\beta}_{\xi}^{*}}^{(j)\T}\bm{A}_\xi^{*(j+1)}\bm{\Delta}_{\bm{\beta}_{\xi}^{*}}^{(j)})^{1/2}}
				{2\bm{\Delta}_{\bm{\beta}_{\xi}^{*}}^{(j)\T}\bm{A}_\xi^{*(j)}\bm{\Delta}_{\bm{\beta}_{\xi}^{*}}^{(j)}}\Bigg]
				\bm{\Delta}_{\bm{\beta}_{\xi}^{*}}^{(j)\T}(\bm{A}_\xi^{*(j+1)}+\xi^{*(j+1)}\bm{I}_p)\bm{\Delta}_{\bm{\beta}_{\xi}^{*}}^{(j)}\text{d}s \notag\\
	&=\bm{\Delta}_{\bm{\beta}_{\xi}^{*}}^{(j)\T}\bm{A}_\xi^{*(j+1)}\bm{\Delta}_{\bm{\beta}_{\xi}^{*}}^{(j)}
			\{2\bm{\Delta}_{\bm{\beta}_{\xi}^{*}}^{(j)\T}\bm{A}_\xi^{*(j)}\bm{\Delta}_{\bm{\beta}_{\xi}^{*}}^{(j)}
				\bm{\Delta}_{\bm{\beta}_{\xi}^{*}}^{(j)\T}(\bm{A}_\xi^{*(j+1)}+\xi^{*(j+1)}\bm{I}_p)\bm{\Delta}_{\bm{\beta}_{\xi}^{*}}^{(j)} \notag \\
			& \ \hphantom{=}
				+\bm{\Delta}_{\bm{\beta}_{\xi}^{*}}^{(j)\T}(\bm{A}_\xi^{*(j+1)}+\xi^{*(j+1)}\bm{I}_p)\bm{\Delta}_{\bm{\beta}_{\xi}^{*}}^{(j)}
				\bm{\Delta}_{\bm{\beta}_{\xi}^{*}}^{(j)\T}\bm{A}_\xi^{*(j)}\bm{\Delta}_{\bm{\beta}_{\xi}^{*}}^{(j)}\}
				(\bm{\Delta}_{\bm{\beta}_{\xi}^{*}}^{(j)\T}\bm{A}_\xi^{*(j)}\bm{\Delta}_{\bm{\beta}_{\xi}^{*}}^{(j)})^2 \notag \\
			& \ \hphantom{=} / \ 
				[\bm{\Delta}_{\bm{\beta}_{\xi}^{*}}^{(j)\T}(\bm{A}_\xi^{*(j)}+\xi^{*(j)}\bm{I}_p)\bm{\Delta}_{\bm{\beta}_{\xi}^{*}}^{(j)}
				\{\bm{\Delta}_{\bm{\beta}_{\xi}^{*}}^{(j)\T}(\bm{A}_\xi^{*(j)}+\xi^{*(j)}\bm{I}_p)\bm{\Delta}_{\bm{\beta}_{\xi}^{*}}^{(j)}
				\bm{\Delta}_{\bm{\beta}_{\xi}^{*}}^{(j)\T}\bm{A}_\xi^{*(j+1)}\bm{\Delta}_{\bm{\beta}_{\xi}^{*}}^{(j)} \notag \\
			& \ \hphantom{=/}
				+\bm{\Delta}_{\bm{\beta}_{\xi}^{*}}^{(j)\T}(\bm{A}_\xi^{*(j+1)}+\xi^{*(j+1)}\bm{I}_p)\bm{\Delta}_{\bm{\beta}_{\xi}^{*}}^{(j)}
				\bm{\Delta}_{\bm{\beta}_{\xi}^{*}}^{(j)\T}\bm{A}_\xi^{*(j)}\bm{\Delta}_{\bm{\beta}_{\xi}^{*}}^{(j)}\}^2] \notag \\
		& \ \hphantom{=}
				+\bm{\Delta}_{\bm{\beta}_{\xi}^{*}}^{(j)\T}(\bm{A}_\xi^{*(j)}+\xi^{*(j)}\bm{I}_p)\bm{\Delta}_{\bm{\beta}_{\xi}^{*}}^{(j)}
				\{2\bm{\Delta}_{\bm{\beta}_{\xi}^{*}}^{(j)\T}(\bm{A}_\xi^{*(j+1)}+\xi^{*(j+1)}\bm{I}_p)\bm{\Delta}_{\bm{\beta}_{\xi}^{*}}^{(j)}
				\bm{\Delta}_{\bm{\beta}_{\xi}^{*}}^{(j)\T}\bm{A}_\xi^{*(j)}\bm{\Delta}_{\bm{\beta}_{\xi}^{*}}^{(j)} \notag \\
			& \ \hphantom{=+}
				+\bm{\Delta}_{\bm{\beta}_{\xi}^{*}}^{(j)\T}(\bm{A}_\xi^{*(j)}+\xi^{*(j)}\bm{I}_p)\bm{\Delta}_{\bm{\beta}_{\xi}^{*}}^{(j)}
				\bm{\Delta}_{\bm{\beta}_{\xi}^{*}}^{(j)\T}\bm{A}_\xi^{*(j+1)}\bm{\Delta}_{\bm{\beta}_{\xi}^{*}}^{(j)}\}
				(\bm{\Delta}_{\bm{\beta}_{\xi}^{*}}^{(j)\T}\bm{A}_\xi^{*(j+1)}\bm{\Delta}_{\bm{\beta}_{\xi}^{*}}^{(j)})^2 \notag \\
			& \ \hphantom{=+}/ \ 
				[\bm{\Delta}_{\bm{\beta}_{\xi}^{*}}^{(j)\T}(\bm{A}_\xi^{*(j+1)}+\xi^{*(j+1)}\bm{I}_p)\bm{\Delta}_{\bm{\beta}_{\xi}^{*}}^{(j)}
				\{\bm{\Delta}_{\bm{\beta}_{\xi}^{*}}^{(j)\T}(\bm{A}_\xi^{*(j)}+\xi^{*(j)}\bm{I}_p)\bm{\Delta}_{\bm{\beta}_{\xi}^{*}}^{(j)}
				\bm{\Delta}_{\bm{\beta}_{\xi}^{*}}^{(j)\T}\bm{A}_\xi^{*(j+1)}\bm{\Delta}_{\bm{\beta}_{\xi}^{*}}^{(j)} \notag \\
			& \ \hphantom{=+/ \ }
				+\bm{\Delta}_{\bm{\beta}_{\xi}^{*}}^{(j)\T}(\bm{A}_\xi^{*(j+1)}+\xi^{*(j+1)}\bm{I}_p)\bm{\Delta}_{\bm{\beta}_{\xi}^{*}}^{(j)}
				\bm{\Delta}_{\bm{\beta}_{\xi}^{*}}^{(j)\T}\bm{A}_\xi^{*(j)}\bm{\Delta}_{\bm{\beta}_{\xi}^{*}}^{(j)}\}^2] \notag \\
	& = \bm{C}(\bm{A}_\xi^{*(j)}, \bm{A}_\xi^{*(j)}+\xi^{*(j)}\bm{I}_p).
	\label{Stryhn}
\end{align}
Here, the second equality holds because of a result in \cite{Str96}. Therefore, from \eqref{SupQ}, \eqref{Qargsup}, \eqref{BaBaS}, and \eqref{Stryhn}, we can obtain the following theorem.

\begin{theorem}
Under the conditions in Theorem \ref{theorem1} and \eqref{condition}, the asymptotic bias in \eqref{zenbias} is given by
	\begin{align}
		&\E\{b_\xi(\bm{k}^*,\bm{\beta}_{\xi}^{*})\}
			=2\sum_{j=1}^m \bm{C}(\bm{A}_\xi^{*(j)}, \bm{A}_\xi^{*(j)}+\xi^{*(j)}\bm{I}_p)
				+\sum_{j=1}^{m+1}{\rm tr}\{\bm{A}_\xi^{*(j)}(\bm{A}_\xi^{*(j)}+\xi^{*(j)}\bm{I}_p)^{-1}\}.
		\label{tt2}
	\end{align}
	\label{theorem2}
\end{theorem}

\noindent
This gives an information criterion as the bias-corrected maximum regularized log-partial likelihood; however, because the asymptotic bias in \eqref{tt2} contains unknown parameters, they are replaced by consistent estimators, as in the TIC and the generalized information criterion (GIC, \citealt{KonKit96}). As a result, for the case where estimation is based on the partial likelihood with the addition of a regularization term in the L$_2$ norm, we propose the following information criterion for the Cox proportional hazards model with change-points:
\begin{align}
	\text{AIC}_\xi=&-2l_\xi(\hat{\bm{\beta}},\hat{\bm{k}}; \bm{t}) 
		+4\sum_{j=1}^m \hat{\bm{C}}\{\hat{\bm A}_\xi^{*(j)}(\hat{\bm{\beta}}, \hat{\bm{k}}), 
			\hat{\bm A}_\xi^{*(j)}(\hat{\bm{\beta}}, \hat{\bm{k}})+\xi^{*(j)}\bm{I}_p\}
\notag \\
 		&+2\sum_{j=1}^{m+1}\text{tr}[\hat{\bm{A}}_\xi^{*(j)}(\hat{\bm{\beta}}, \hat{\bm{k}})
			\{\hat{\bm{A}}_\xi^{*(j)}(\hat{\bm{\beta}}, \hat{\bm{k}})+\xi^{*(j)}\bm{I}_{p}\}^{-1}],
	\label{henkaAICKappa}
\end{align}
where $\hat{\bm{C}}(\bm{A}^{(j)\dagger}, \bm{A}^{(j)\ddagger})$ is $\bm{C}(\bm{A}^{(j)\dagger}, \bm{A}^{(j)\ddagger})$ with $\bm{\beta}_{\xi}^{*}$ replaced by $\hat{\bm{\beta}}$, and $\hat{\bm{A}}_\xi^{*(j)}$ is the $p\times p$ matrix defined in \eqref{defhatA}.

Although so far we have discussed the model given by \eqref{CoxRegCP}, in which all the regression parameters are structurally changed at change-points, even in the change-point model that some of the parameters are structurally changed as follows
\begin{align*}
	\lambda(t\mid \bm{z})=\lambda_0(t)\exp(\bm{\beta}_1^{(j)\T}\bm{\bm{z}}_1+\bm{\beta}_2^{\T}\bm{\bm{z}}_2),
\end{align*}
the asymptotic bias is derived similarly to that given in \eqref{tt2} under the same conditions, and we can obtain the same AIC as that given in \eqref{henkaAICKappa}. Also, when the regularization parameter $\xi$ is 0, $\bm{\beta}_{\xi}^{*}$ becomes equal to $\bm{\beta}^{*}$; then, by using $\bm{C}(\bm{A}_\xi^{*(j)},\bm{A}_\xi^{*(j)})=\hat{\bm{C}}\{\hat{\bm{A}}_\xi^{*(j)}(\hat{\bm{\beta}},\hat{\bm{k}}),\hat{\bm{A}}_\xi^{*(j)}(\hat{\bm{\beta}}, \hat{\bm{k}})\}=3/2$, we obtain the following corollary.

\begin{corollary}
Under the conditions in Theorem \ref{theorem1} and \eqref{condition}, the asymptotic bias given in \eqref{zenbias} and based on the conventional partial likelihood method, which uses \eqref{rpll} with $\xi=0$, is given by
	\begin{align*}
		\E\{b(\bm{k}^*,\bm{\beta}^{*})\}=3m+p(m+1).
	\end{align*}
	\label{corollary1}
\end{corollary}

\noindent
From this, we can see that the asymptotic bias due to the change-point parameter is three times greater than that due to the regression parameter, which is consistent with the result in \cite{Nin15}.
As a result, we propose the following criterion:
\begin{align}
	\text{AIC}=-2l(\hat{\bm{\beta}},\hat{\bm{k}}; \bm{t})
		+6m+2p(m+1),
	\label{henkaAIC}
\end{align}
which we call the AIC for the Cox proportional hazards model with change-points, for estimation based on the conventional partial likelihood method. 

\section{Numerical experiments}
\label{sec4}
In this section, we use the results of numerical experiments to examine the performance of the proposed AIC given in \eqref{henkaAIC} (hereafter referred to simply as ``AIC'') as an information criterion for estimation based on the conventional partial likelihood method without regularization. For comparison, we also consider the following information criterion:
\begin{align*}
	\text{AIC}_\text{naive}=-2l(\hat{\bm{\beta}},\hat{\bm{k}}; \bm{t})+2m+2p(m+1),
\end{align*}
which handles the bias due to the change-point parameter in the same way that it handles the bias due to the regression parameter. To address the simplest setting, we assume that
\begin{align}
	\label{SimuModel}
	\lambda(t\mid \bm{z})=\bigg\{
	\begin{array}{l}
		\lambda_0(t)\exp({\beta}^{(1)}{z}), \qquad t\in[0,k)\\
		\lambda_0(t)\exp({\beta}^{(2)}{z}), \qquad t\in[k,T)
	\end{array}
\end{align}
gives a univariate Cox proportional hazards model with one change-point $k$. As this experimental model has one change-point parameter and two regression parameters, the asymptotic bias evaluations for AIC and $\text{AIC}_\text{naive}$ are $3\times1+1\times2=5$ and $1\times1+1\times2=3$, respectively.

First, to examine whether these penalty terms provide accurate approximations of the bias in the maximum log-partial likelihood, we numerically evaluated the bias with different true parameter values and different data sizes in the model given by \eqref{SimuModel}. The results are listed in Table \ref{tBias}. In every setting, the value was around $5$, and a value of at least $5$ was a more accurate approximation of the bias than $3$. These results indicate that AIC is a more accurate approximation of the Kullback-Leibler divergence than $\text{AIC}_\text{naive}$ is.

\begin{table}[t!]
\caption{Bias in the maximum log-partial likelihood. The values are means (standard errors in parentheses) obtained by a Monte Carlo method through 100 iterations based on the model given by \eqref{SimuModel}. The true change-point is the point at which the true survival probability has reached $100\times(1-\alpha)\%$.}
\begin{center}
\begin{tabular*}{0.95\textwidth}{@{\extracolsep{\fill}}ccccccc}
\cmidrule(r){1-7}
$\alpha$ & $\exp(\bm{\beta}^{*(1)})$ & $\exp(\bm{\beta}^{*(2)})$ & $\# D$: 50 & $\# D$: 100 & $\# D$: 150 & $\# D$: 200 \\
\cmidrule(r){1-7}
 & & 0.9 & 4.21 (0.35) & 5.73 (0.41) & 6.75 (0.44) & 5.53 (0.43) \\
\multirow{2}{*}{0.3} & \multirow{2}{*}{1.0} & 0.8 & 4.76 (0.43) & 5.33 (0.45) & 6.12 (0.47) & 5.51 (0.45) \\
 & & 0.7 & 4.81 (0.45) & 5.86 (0.56) & 5.75 (0.46) & 5.57 (0.44) \\
 & & 0.6 & 4.73 (0.48) & 5.16 (0.45) & 6.65 (0.58) & 5.46 (0.49) \\
\cmidrule(r){1-7}
 & & 0.9 & 4.78 (0.47) & 5.16 (0.44) & 5.54 (0.50) & 5.41 (0.49) \\
\multirow{2}{*}{0.5} & \multirow{2}{*}{1.0} & 0.8 & 4.99 (0.49) & 5.25 (0.47) & 5.44 (0.49) & 5.25 (0.43) \\
 & & 0.7 & 4.69 (0.42) & 5.47 (0.47) & 5.92 (0.54) & 5.34 (0.37) \\
 & & 0.6 & 4.51 (0.43) & 5.23 (0.46) & 6.11 (0.54) & 5.26 (0.41) \\
\cmidrule(r){1-7}
 & & 0.9 & 4.89 (0.38) & 5.93 (0.67) & 5.45 (0.41) & 5.97 (0.51) \\
\multirow{2}{*}{0.7} & \multirow{2}{*}{1.0} & 0.8 & 4.83 (0.32) & 6.03 (0.69) & 5.43 (0.40) & 6.25 (0.50) \\
 & & 0.7 & 4.93 (0.37) & 5.99 (0.68) & 5.56 (0.43) & 6.30 (0.58) \\
 & & 0.6 & 4.67 (0.40) & 5.79 (0.67) & 5.45 (0.41) & 5.10 (0.48) \\
\cmidrule(r){1-7}
\end{tabular*}
\end{center}
\label{tBias}
\end{table}

Second, to actually compare the performances of AIC and $\text{AIC}_{\text{naive}}$, we considered models given by
\begin{align}
	\label{SimuModel2}
	\lambda(t\mid {z})=\lambda_0(t)\exp({\beta}^{(j)}{z}), \qquad t\in[k^{(j-1)},k^{(j)}), \qquad j\in\{1,2,\ldots,m+1\},
\end{align}
with $m=0$, $m=1$, $m=2$, and $m=3$. Then, we selected the optimal model for each criterion. Here, $k^{(0)}=0$ and $k^{(m+1)}=T$. Table \ref{t01rate} summarizes the Kullback-Leibler divergence between the true and estimated distributions. It also gives the selection probabilities under the true structure determined under a setting in which the number of change-points was 0 or 1, with different true values for the change-point parameter and the amount of changes.

For the case of no change-points, i.e., $m^*=0$, we can see that, regardless of the event size, AIC could select the model with no change-point (i.e., with $m=0$) with a high probability of approximately 90\% or higher. On the other hand, $\text{AIC}_{\text{naive}}$ selected the model with change-points (i.e., the model with $m>0$) with a probability of approximately 50\% or higher, and this trend was even more apparent when the event size was large. This result implies that $\text{AIC}_{\text{naive}}$ underestimates the asymptotic bias and causes overfitting. Moreover, for the case of one true change-point, when the event size and the amount of change were smaller, AIC was less likely than $\text{AIC}_{\text{naive}}$ to select the number of true change-points. However, AIC gave a clearly smaller Kullback-Leibler divergence than $\text{AIC}_{\text{naive}}$ under any setting, and we can thus say that AIC clearly selects the better model in terms of prediction.

\begin{table}[p]
\caption{Kullback-Leibler divergence (K-L) between the true and estimated distributions, and the probability of selecting 0, 1, 2, or 3 change-points ($\%$). These values were obtained by a Monte Carlo method through 100 iterations with fixed true parameters based on the model given by \eqref{SimuModel}.}
\begin{center}
\begin{tabular*}{0.95\textwidth}{@{\extracolsep{\fill}}cccccccrrrr}
\cmidrule(r){1-11}
$\alpha$ & $\# D$ & $\exp(\bm{\beta}^{*(1)})$ & $\exp(\bm{\beta}^{*(2)})$ & $m^*$ & & \multicolumn{1}{c}{K-L} & \multicolumn{1}{c}{0 ($\%$)} & \multicolumn{1}{c}{1 ($\%$)} & \multicolumn{1}{c}{2 ($\%$)} & \multicolumn{1}{c}{3 ($\%$)} \\
\cmidrule(r){1-11}
 & & & 1.00 & 0 & $\text{AIC}_{\text{naive}}$ & 2.99 & 58 & 18 & 16 & 8 \\
 & & & & & AIC & 0.78 & 95 & 4 & 1 & 0 \\
\multirow{2}{*}{0.3} & \multirow{2}{*}{50} & \multirow{2}{*}{1.00} & 0.50 & 1 & $\text{AIC}_{\text{naive}}$ & 2.91 & 40 & 34 & 20 & 6 \\
 & & & & & AIC & 0.91 & 89 & 11 & 0 & 0 \\
 & & & 0.25 & 1 & $\text{AIC}_{\text{naive}}$ & 6.48 & 12 & 56 & 25 & 7 \\
 & & & & & AIC & 3.58 & 53 & 43 & 4 & 0 \\
\cmidrule(r){1-11}
 & & & 1.00 & 0 & $\text{AIC}_{\text{naive}}$ & 4.92 & 41 & 16 & 19 & 24 \\
 & & & & & AIC & 1.26 & 88 & 10 & 2 & 0 \\
\multirow{2}{*}{0.3} & \multirow{2}{*}{100} & \multirow{2}{*}{1.00} & 0.50 & 1 & $\text{AIC}_{\text{naive}}$ & 4.72 & 23 & 20 & 28 & 29 \\
 & & & & & AIC & 1.33 & 69 & 28 & 3 & 0 \\
 & & & 0.25 & 1 & $\text{AIC}_{\text{naive}}$ & 5.37 & 1 & 41 & 26 & 32 \\
 & & & & & AIC & 1.69 & 26 & 67 & 6 & 1 \\
\cmidrule(r){1-11}
 & & & 1.00 & 0 & $\text{AIC}_{\text{naive}}$ & 3.73 & 48 & 25 & 17 & 10 \\
 & & & & & AIC & 1.15 & 92 & 7 & 1 & 0 \\
\multirow{2}{*}{0.5} & \multirow{2}{*}{50} & \multirow{2}{*}{1.00} & 0.50 & 1 & $\text{AIC}_{\text{naive}}$ & 2.89 & 39 & 33 & 18 & 10 \\
 & & & & & AIC & 1.06 & 83 & 17 & 0 & 0 \\
 & & & 0.25 & 1 & $\text{AIC}_{\text{naive}}$ & 4.58 & 6 & 58 & 31 & 5 \\
 & & & & & AIC & 2.70 & 47 & 53 & 0 & 0 \\
\cmidrule(r){1-11}
 & & & 1.00 & 0 & $\text{AIC}_{\text{naive}}$ & 4.66 & 40 & 16 & 23 & 21 \\
 & & & & & AIC & 1.16 & 91 & 6 & 2 & 1 \\
\multirow{2}{*}{0.5} & \multirow{2}{*}{100} & \multirow{2}{*}{1.00} & 0.50 & 1 & $\text{AIC}_{\text{naive}}$ & 4.82 & 9 & 29 & 35 & 27 \\
 & & & & & AIC & 1.96 & 63 & 33 & 3 & 1 \\
 & & & 0.25 & 1 & $\text{AIC}_{\text{naive}}$ & 4.68 & 0 & 39 & 22 & 39 \\
 & & & & & AIC & 2.02 & 7 & 82 & 9 & 2 \\
\cmidrule(r){1-11}
 & & & 1.00 & 0 & $\text{AIC}_{\text{naive}}$ & 2.85 & 55 & 24 & 12 & 9 \\
 & & & & & AIC & 0.86 & 93 & 7 & 0 & 0 \\
\multirow{2}{*}{0.7} & \multirow{2}{*}{50} & \multirow{2}{*}{1.00} & 0.50 & 1 & $\text{AIC}_{\text{naive}}$ & 2.62 & 41 & 37 & 14 & 8 \\
 & & & & & AIC & 0.90 & 89 & 11 & 0 & 0 \\
 & & & 0.25 & 1 & $\text{AIC}_{\text{naive}}$ & 3.15 & 13 & 49 & 25 & 13 \\
 & & & & & AIC & 1.75 & 59 & 36 & 5 & 0 \\
\cmidrule(r){1-11}
 & & & 1.00 & 0 & $\text{AIC}_{\text{naive}}$ & 4.76 & 33 & 21 & 28 & 18 \\
 & & & & & AIC & 0.96 & 93 & 6 & 1 & 0 \\
\multirow{2}{*}{0.7} & \multirow{2}{*}{100} & \multirow{2}{*}{1.00} & 0.50 & 1 & $\text{AIC}_{\text{naive}}$ & 4.25 & 15 & 33 & 27 & 25 \\
 & & & & & AIC & 1.26 & 71 & 27 & 2 & 0 \\
 & & & 0.25 & 1 & $\text{AIC}_{\text{naive}}$ & 4.05 & 4 & 37 & 30 & 29 \\
 & & & & & AIC & 2.19 & 22 & 72 & 4 & 2 \\
\cmidrule(r){1-11}
\end{tabular*}
\end{center}
\label{t01rate}
\end{table}

Finally, we compared the performance in a more practical setting. In the model given by \eqref{SimuModel2}, we assumed that the number of true change-points was $1$ and the true survival probability was $100\times(1-\alpha)\%$, for a random variable $\alpha$ following a continuous uniform distribution over $[0.1, 0.9]$. We further assumed that the true amounts $\exp({\beta}^{*(2)})/\exp({\beta}^{*(1)})$ were given by $2^{u_1(\psi+u_2)}$, where $u_1$ and $u_2$ were independent random variables that were distributed according to a discrete uniform distribution over $\{-1,1\}$ and a continuous uniform distribution over $[0,1]$, respectively. For the true structure determined randomly in this way, Table \ref{t02rate} summarizes the Kullback-Leibler divergence between the true and selected models and the probability of selecting each model. Similarly to the results in Table \ref{t01rate}, AIC gave a smaller Kullback-Leibler divergence than $\text{AIC}_{\text{naive}}$ under any setting, and we can thus say that AIC is the better information criterion in terms of prediction. In particular, when the event size was large, $\text{AIC}_{\text{naive}}$ tended to select too many change-points, and we suggest that this is one reason why its Kullback-Leibler divergence values were large.

\begin{table}[t!]
\caption{Kullback-Leibler divergence (K-L) between the true and estimated distributions, and the probability of selecting 0, 1, 2, or 3 change-points ($\%$). These values were obtained by a Monte Carlo method through 100 iterations with varying true parameters based on the model given by \eqref{SimuModel}.}
\begin{center}
\begin{tabular*}{0.95\textwidth}{@{\extracolsep{\fill}}cccccrrrr}
\cmidrule(r){1-9}
$\# D$ & $\exp(\bm{\beta}^{*(1)})$ & $\psi$ & & K-L & \multicolumn{1}{c}{0 ($\%$)} & \multicolumn{1}{c}{1 ($\%$)} & \multicolumn{1}{c}{2 ($\%$)} & \multicolumn{1}{c}{3 ($\%$)} \\
\cmidrule(r){1-9}
 & & 1.00 & $\text{AIC}_{\text{naive}}$ & 4.16 & 20 & 41 & 25 & 14 \\
 & & & AIC & 2.01 & 67 & 31 & 2 & 0 \\
 & & 1.50 & $\text{AIC}_{\text{naive}}$ & 3.29 & 24 & 43 & 25 & 8 \\
\multirow{2}{*}{50} & \multirow{2}{*}{1.00} & & AIC & 2.45 & 64 & 35 & 1 & 0 \\
 & & 2.00 & $\text{AIC}_{\text{naive}}$ & 5.18 & 13 & 65 & 18 & 4 \\
 & & & AIC & 4.13 & 44 & 51 & 5 & 0 \\
 & & 2.50 & $\text{AIC}_{\text{naive}}$ & 7.64 & 8 & 56 & 23 & 13 \\
 & & & AIC & 4.58 & 25 & 72 & 3 & 0 \\
\cmidrule(r){1-9}
 & & 0.50 & $\text{AIC}_{\text{naive}}$ & 4.23 & 26 & 20 & 28 & 26 \\
 & & & AIC & 1.07 & 67 & 31 & 2 & 0 \\
 & & 1.00 & $\text{AIC}_{\text{naive}}$ & 4.41 & 8 & 28 & 33 & 31 \\
\multirow{2}{*}{100} & \multirow{2}{*}{1.00} & & AIC & 1.30 & 50 & 45 & 5 & 0 \\
 & & 1.50 & $\text{AIC}_{\text{naive}}$ & 6.06 & 9 & 31 & 27 & 33 \\
 & & & AIC & 1.37 & 38 & 57 & 4 & 1 \\
 & & 2.00 & $\text{AIC}_{\text{naive}}$ & 7.03 & 1 & 31 & 32 & 36 \\
 & & & AIC & 2.55 & 12 & 80 & 6 & 2 \\
\cmidrule(r){1-9}
 & & 0.25 & $\text{AIC}_{\text{naive}}$ & 6.38 & 10 & 15 & 26 & 49 \\
 & & & AIC & 1.69 & 66 & 24 & 7 & 3 \\
 & & 0.50 & $\text{AIC}_{\text{naive}}$ & 6.57 & 6 & 14 & 24 & 56 \\
\multirow{2}{*}{200} & \multirow{2}{*}{1.00} & & AIC & 1.96 & 45 & 48 & 5 & 2 \\
 & & 1.00 & $\text{AIC}_{\text{naive}}$ & 6.61 & 1 & 13 & 24 & 62 \\
 & & & AIC & 2.34 & 23 & 66 & 8 & 3 \\
 & & 1.50 & $\text{AIC}_{\text{naive}}$ & 6.59 & 1 & 16 & 40 & 43 \\
 & & & AIC & 2.47 & 12 & 72 & 13 & 3 \\
\cmidrule(r){1-9}
\end{tabular*}
\end{center}
\label{t02rate}
\end{table}

\section{Real data analysis}
\label{sec5}

In this section, we apply the AIC in \eqref{henkaAIC} and $\text{AIC}_{\text{naive}}$ to data from a randomized, placebo-controlled clinical trial of patients with malignant glioma. The clinical trial was designed to examine the effects of a biodegradable polymer that contained carmustine and was implanted into a brain tumor site after surgical resection of recurrent tumors. The clinical trial also sought to examine whether the carmustine-impregnated polymer could provide more sustained local exposure to chemotherapeutic agents that prolong survival. A total of 222 patients were enrolled from 27 institutions: 110 patients were randomly assigned to the test group, while the other 112 were assigned to the control group. The clinical trial design and analysis results were reported in \cite{brem1995placebo}. We inferred that the survival curves of the two groups, categorized by whether 75\% or more of the tumor was resected in the clinical trial, would diverge after a certain period after resection.

\begin{figure}[t!]
\vspace{4mm}
\begin{center}
\includegraphics[keepaspectratio,width=0.95\textwidth]
 {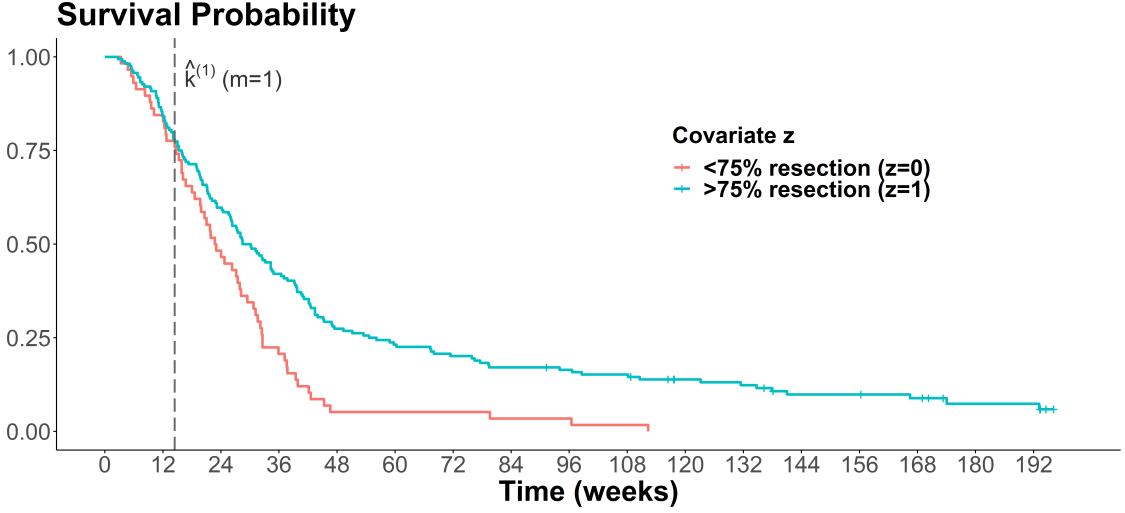}
\end{center}
\vspace{-4mm}
\caption{Kaplan-Meier curves for the $<$75\% and $>$75\% resection groups. The upper limit for the change-point is set at 48 weeks because approximately 95\% of events in the group with $<$75\% resection occurred by 48 weeks, with the remaining events occurring after approximately 80 weeks.}
 \label{fRealData}
\end{figure}

Then, we searched for change-points by applying AIC and ${\rm AIC}_{\rm naive}$ to the data created by weighting each individual by 2 to check the behavior under a certain number of events. Specifically, letting $\bm{z}$ be the variable that indicates whether 75\% or more of the tumor is resected, we applied the two information criteria to select the optimal model among the models given by \eqref{SimuModel} with $m=0,1,2,3$. The results are listed in Table \ref{tRealData}. Whereas ${\rm AIC}_{\rm naive}$ selected a model with three change-points, at 8.3, 10.6, and 14.4 weeks, AIC selected a model with one change-point, at 14.4 weeks. This could suggest that ${\rm AIC}_{\rm naive}$ selected a complex model because it underestimated the asymptotic bias. Figure \ref{fRealData} shows Kaplan-Meier curves for the $<$75\% and $>$75\% resection groups. The two curves overlapped for less than 16 weeks, and then the difference between the curves increased, which makes it reasonable to expect that a structural change occurred around that time.

\begin{table}[t!]
\caption{Change-point estimates $\hat{\bm{k}}$, maximum log-partial likelihood $l(\hat{\bm{\beta}},\hat{\bm{k}};\bm{t})$, AIC, and $\text{AIC}_\text{naive}$ obtained from real clinical trial data.}
\begin{center}
\begin{tabular*}{0.95\textwidth}{@{\extracolsep{\fill}}ccrrrccc}
\cmidrule(r){1-8}
$m$ &
 $p(m+1)$ & 
 \multicolumn{1}{c}{$\hat{k}^{(1)}$} &
 \multicolumn{1}{c}{$\hat{k}^{(2)}$} &
 \multicolumn{1}{c}{$\hat{k}^{(3)}$} &
 $l(\hat{\bm{\beta}},\hat{\bm{k}};\bm{t})$ &
 $\text{AIC}_\text{naive}$ &
 AIC \\ 
\cmidrule(r){1-8}
0 & 1 & & & & $-$2169.65 & 4341.29 & 4341.29 \\
1 & 2 & 14.4 & & & $-$2164.92 & 4335.83 & 4339.83 \\
2 & 3 & 10.6 & 14.4 & & $-$2161.72 & 4333.44 & 4341.44 \\
3 & 4 & 8.3 & 10.6 & 14.4 & $-$2158.79 & 4331.59 & 4343.59 \\ \cmidrule(r){1-8}
\end{tabular*}
\end{center}
\label{tRealData}
\end{table}

\section{Extension}
\label{sec6}
In this section, under the setting of $\xi=0$, we extend the AIC in \eqref{henkaAIC} to allow for model misspecification. As before, we defined $\bm{\beta}^*=(\bm{\beta}^{*(1)\T},\bm{\beta}^{*(2)\T},\ldots,\bm{\beta}^{*(m+1)\T})^{\T}$ as $\argsup_{\bm{\beta}}\E\{l(\bm{\beta},\bm{k}^*;\allowbreak\bm{t})\}$. For the case without model misspecification, it denotes the true value of the regression parameter vector $\bm{\beta}=(\bm{\beta}^{(1)\T},\bm{\beta}^{(2)\T},\ldots,\bm{\beta}^{(m+1)\T})^{\T}$ for the model given by \eqref{CoxRegCP}. However, this model potentially assumes log-linearity for the relationship between the covariate $\bm{z}$ and the hazard function $\lambda(t\mid\bm{z})$. Accordingly, application of this model to a situation in which this assumption does not hold would cause the model to be misspecified. Moreover, the model is misspecified in two cases: when the covariates that can be included in it are restricted, and when the conditional independence, given $\bm{z}$, between the occurrence times of events and censoring, $y_1$ and $y_2$, does not hold. Hence, we derive an information criterion for the model given by \eqref{CoxRegCP} for the case of model misspecification, where $\E^*$ denotes the expectation with respect to the distribution following the unknown model. Note that $\bm{\beta}^*$ is not necessarily the true value for the case of model misspecification. 

From \cite{struthers1986misspecified} and \cite{lin1989robust}, the following holds:
\begin{align*}
	&\sqrt{n}(\hat{\bm{\beta}}_{\bm{k}^*}^{(j)}-\bm{\beta}^{*(j)})
		\stackrel{\rm d}{\to}
		{\rm N}(\bm{0}_p, \bm{B}_0^{*(j)-1}\bm{A}_0^{*(j)}\bm{B}_0^{*(j)-1}).
\end{align*}
Here, by letting
\begin{align*}
	&\bm{w}^{(j)}(\bm{\beta},\bm{k}) 
	\equiv \sum_{i\in D^{(j)}}\{\bm{z}_i-\bm{h}(t_i,\bm{\beta}^{(j)})\}
		-\sum_{i=1}^{n}\sum_{l\in D^{(j)}:t_l<t_i}
			\frac{\exp(\bm{\beta}^{(j)\T}\bm{z}_i)}{\sum_{i'\in R(t_l)}\exp(\bm{\beta}^{(j)\T}\bm{z}_{i'})}
			\{\bm{z}_i-\bm{h}(t_l,\bm{\beta}^{(j)})\},
\end{align*}
we define
\begin{align*}
\bm{A}_0^{*(j)}\equiv\E^*\bigg\{\frac{1}{n}\bm{w}^{(j)}(\bm{\beta}^*,\bm{k}^*)\bm{w}^{(j)}(\bm{\beta}^*,\bm{k}^*)^\T\bigg\}.
\end{align*}
We also specify that $\bm{B}_0^{*(j)}$ is $\bm{B}_{\xi}^{(j)}(\bm{\beta}^*, \bm{k}^*)$ with $\xi=0$. 

Let $k^{(j)}=k^{*(j)}+s^{(j)}/n$ for each $j\in\{1, 2, \ldots, m\}$. First, we consider the case where $\bm{s}=(s^{(1)},s^{(2)},\ldots,s^{(m)})^{\T}$ is a vector with finite values. As with \eqref{KOrderInBeta}, it can be shown that
\begin{align}
	\hat{\bm{\beta}}_{\bm{k}}-\hat{\bm{\beta}}_{\bm{k}^*}=\OP(1/n).
	\label{KOrderInBeta_TIC}
\end{align}
Then, by the same reasoning as for \eqref{AsyNorm}, it follows that
\begin{align}
	&\sqrt{n}(\hat{\bm{\beta}}_{\bm{k}}^{(j)}-\bm{\beta}^{*(j)})
		\stackrel{\rm d}{\to}
		{\rm N}(\bm{0}_p, \bm{B}_0^{*(j)-1}\bm{A}_0^{*(j)}\bm{B}_0^{*(j)-1}),
		\label{AsyNorm_TIC}
\end{align}
and by using a two-sided random walk with a negative drift $Q_{k,\bm{t}}^{(j)}$ defined by $Q_{\xi; k,\bm{t}}^{(j)}$ with $\xi=0$, we obtain
\begin{align}\label{llTaylorBetaHatKBetaHatKStar_TIC}
	l(\hat{\bm{\beta}}_{\bm{k}},\bm{k}; \bm{t})-l(\hat{\bm{\beta}}_{\bm{k}^*},\bm{k}^*; \bm{t}) 
		&=l(\bm{\beta}_{\xi}^{*},\bm{k}; \bm{t})-l(\bm{\beta}_{\xi}^{*},\bm{k}^*; \bm{t})+\oP(1) \notag \\
		&=\sum_{j=1}^m Q_{k^{*(j)}+s^{(j)}/n, \bm{t}}^{(j)}+\oP(1)=\OP(1).
\end{align}
Furthermore, from Taylor expansion around $\hat{\bm{\beta}}_{\bm{k}^*}^{(j)}=\bm{\beta}^{*(j)}$ for the log-partial likelihood, \eqref{AsyNorm_TIC}, and \cite{Murphy2000profile}, we can show that
\begin{align}\label{llTaylorBetaHatKStarBetaStarKStar_TIC}
	l(\hat{\bm{\beta}}_{\bm{k}^*},\bm{k}^*; \bm{t})-l(\bm{\beta}^*,\bm{k}^*; \bm{t})
		&=\frac{1}{2}\sum_{j=1}^{m+1}\bm{\nu}^{(j)\T}\bm{\nu}^{(j)}+\oP(1)=\OP(1),
\end{align}
where $\bm{\nu}^{(j)}$ is a random vector distributed according to a multivariate normal distribution ${\rm N}(\bm{0}_{p}, \bm{B}_0^{*(j)-1}\bm{A}_0^{*(j)})$. Therefore, from \eqref{llTaylorBetaHatKBetaHatKStar_TIC} and \eqref{llTaylorBetaHatKStarBetaStarKStar_TIC}, $\hat{ll}(\bm{k}; \bm{t},\bm{t})=\OP(1)$ holds.

Next, we consider the case where $\bm{s}$ is not a vector with finite values. Again, as in Section \ref{sec3}, let us consider only the following situation:
\begin{equation*}
	\bigg\{
	\begin{array}{l}
		k^{(j')}=k^{*(j')}+s^{(j')}/n, \qquad 0>s^{(j')}\neq\O(1)\\
		k^{(j)}=k^{*(j)}+s^{(j)}/n, \qquad s^{(j)}=\O(1) \qquad (j\neq j').
	\end{array}
\end{equation*}
In this case, if $\xi=0$ in \eqref{llRatioBetaHatKBetaHatKStar}, then it follows that $\P\{l(\hat{\bm{\beta}}_{\bm{k}},\bm{k})-l(\hat{\bm{\beta}}_{\bm{k}^*},\bm{k}^*)>-M\}\to0$ for any $M>0$, and from \eqref{llTaylorBetaHatKStarBetaStarKStar_TIC}, we have $\P\{\hat{ll}(\bm{k}; \bm{t},\bm{t})>-M\}\to0$ for any $M>0$. Therefore, $\hat{\bm{k}}-\bm{k}^*=\OP(1/n)$ holds.

From the above derivation, we obtain $K=\{\bm{k}\mid k^{(j)}\in K^{(j)},\ j\in\{1,2,\ldots,m\}\}$, where $K^{(j)}=\{k\mid k-k^{*(j)}=\O(1/n)\}$. Then, from \eqref{llTaylorBetaHatKBetaHatKStar_TIC} and \eqref{llTaylorBetaHatKStarBetaStarKStar_TIC}, we have
\begin{align}
	&\sup_{\bm{k}\in K}\hat{ll}(\bm{k}; \bm{t},\bm{t})
		=\sum_{j=1}^m \sup_{k\in K^{(j)}}Q_{k,\bm{t}}^{(j)}+\frac{1}{2}\sum_{j=1}^{m+1}\bm{\nu}^{(j)\T}\bm{\nu}^{(j)}+\oP(1) 
	\label{SupllHat_TIC}
\end{align}
and
\begin{align}
	&\argsup_{\bm{k}\in K}\hat{ll}(\bm{k}; \bm{u},\bm{u})
		=\bigg(\argsup_{k\in K^{(1)}}Q_{k,\bm{u}}^{(1)}, \argsup_{k\in K^{(2)}}Q_{k,\bm{u}}^{(2)},\ldots,
		\argsup_{k\in K^{(m)}}Q_{k,\bm{u}}^{(m)}\bigg)^\T+\oP(1).
	\label{ArgsupllHat_TIC}
\end{align}
In addition, by letting $\check{k}_{\bm{u}}^{(j)}\equiv\argsup_{k\in K^{(j)}}Q_{k,\bm{u}}^{(j)}$ and $\check{\bm{k}}_{\bm{u}}\equiv(\check{k}_{\bm{u}}^{(1)},\check{k}_{\bm{u}}^{(2)},\ldots,\check{k}_{\bm{u}}^{(m)})^\T$, it follows that
\begin{align}
	\hat{\bm{\beta}}_{\check{\bm{k}}_{\bm{u}},\bm{u}}-\hat{\bm{\beta}}_{\bm{k}^*,\bm{u}}=\OP(1/n)
	\label{BetaHatTildeKBetaHatKStar_TIC}
\end{align}
and
\begin{align}
	\hat{\bm{\beta}}_{\check{\bm{k}}_{\bm{u}},\bm{u}}-\bm{\beta}^{*}=\OP(1/\sqrt{n})	
	\label{BetaHatTildeKBetaStar_TIC}.
\end{align}
For $\hat{ll}(\check{\bm{k}}_{\bm{u}}; \bm{t},\bm{u})$, from \cite{Murphy2000profile}, \eqref{BetaHatTildeKBetaHatKStar_TIC}, and \eqref{BetaHatTildeKBetaStar_TIC} by the same reasoning as for \eqref{llTaylorBetaStarKTildeBetaHatKTilde}, we can obtain
\begin{align}
	\label{llTaylorBetaStarKTildeBetaHatKTilde_TIC}
	&l(\bm{\beta}^{*},\check{\bm{k}}_{\bm{u}}; \bm{t})
		-l(\hat{\bm{\beta}}_{\check{\bm{k}}_{\bm{u}},\bm{u}},\check{\bm{k}}_{\bm{u}}; \bm{t}) \notag \\
	&=-\sum_{j=1}^{m+1}\Bigg[(\hat{\bm{\beta}}_{\bm{k}^*,\bm{u}}^{(j)}-\bm{\beta}^{*(j)})^{\T}
		\sum_{i\in D^{*(j)}}\{\bm{z}_i-\bm{h}(t_i,\bm{\beta}^{*(j)})\}+\frac{1}{2}\bm{\nu}^{(j)\T}\bm{\nu}^{(j)}\Bigg]+\oP(1).
\end{align}
Furthermore, from
\begin{align}
	\label{llTaylorBetaStarKStarBetaStarKTilde_TIC}
	l(\bm{\beta}^{*},\bm{k}^*; \bm{t})-l(\bm{\beta}^{*},\check{\bm{k}}_{\bm{u}}; \bm{t})
		=\sum_{j=1}^m Q_{\check{k}_{\bm{u}}^{(j)},\bm{t}}^{(j)},
\end{align}
and by using \eqref{ArgsupllHat_TIC} and \eqref{llTaylorBetaStarKTildeBetaHatKTilde_TIC}, it follows that
\begin{align}
	\label{llHatKTilde_TIC}
	&\hat{ll}\bigg\{\argsup_{\bm{k}\in K}\hat{ll}(\bm{k}; \bm{u},\bm{u}); \bm{t},\bm{u}\bigg\} \notag \\
	&=\hat{ll}(\check{\bm{k}}_{\bm{u}}; \bm{t},\bm{u})+\oP(1) \notag \\
	&=-\sum_{j=1}^m Q_{\check{k}_{\bm{u}}^{(j)},\bm{t}}^{(j)}
		+\sum_{j=1}^{m+1}\Bigg[(\hat{\bm{\beta}}_{\bm{k}^*,\bm{u}}^{(j)}-\bm{\beta}^{*(j)})^{\T}
		\sum_{i\in D^{*(j)}}\{\bm{z}_i-\bm{h}(t_i,\bm{\beta}^{*(j)})\}-\frac{1}{2}\bm{\nu}^{(j)\T}\bm{\nu}^{(j)}\Bigg]+\oP(1).
\end{align}
Thus, from \eqref{SupllHat_TIC} and \eqref{llHatKTilde_TIC}, we obtain the following corollary.

\begin{corollary}
Under the condition in Theorem \ref{theorem1}, even if model misspecification with $\xi=0$ in \eqref{rpll} exists, the asymptotic bias in \eqref{zenbias} is given by
	\begin{align}
		\E\{b(\bm{k}^*,\bm{\beta}^*)\}=&\sum_{j=1}^m\E\bigg(\sup_{k\in K^{(j)}}Q_{k,\bm{t}}^{(j)}
			+Q_{\argsup_{k\in K^{(j)}}Q_{k,\bm{u}}^{(j)},\bm{t}}^{(j)}\bigg) + \sum_{j=1}^{m+1}\text{tr}(\bm{A}_0^{*(j)}\bm{B}_0^{*(j)-1}).
\notag 
	\end{align}
	\label{corollary2}
\end{corollary}

\noindent
As in Section \ref{sec3}, we assume the following condition:
\begin{align}
	\bm{\beta}^{*(j+1)}-\bm{\beta}^{*(j)}=\bm{\Delta}_{\bm{\beta}^{*}}^{(j)}/\sqrt{\alpha_n} 
	\qquad (j\in\{1,2,\ldots,m\}), \qquad \O(1)\neq\alpha_n=\o(n).
	\label{condition_TIC}
\end{align}
Under this condition, we investigate the asymptotic behavior of the change-point estimator. Similarly to the derivation of Corollary \ref{corollary2}, when $\bm{s}$ is a vector with finite values, we have $\hat{\bm{\beta}}_{\bm{k}}-\hat{\bm{\beta}}_{\bm{k}^*}=\OP(\sqrt{\alpha_n}/n)$ instead of \eqref{KOrderInBeta_TIC}. Then, because \eqref{llTaylorBetaHatKBetaHatKStar_TIC} and \eqref{llTaylorBetaHatKStarBetaStarKStar_TIC} hold, we can show that $\hat{ll}(\bm{k}; \bm{t},\bm{t})=\OP(1)$. On the other hand, when $\bm{s}$ is not a vector with finite values, and again similarly to the derivation of Corollary \ref{corollary2}, let us consider the case where $k^{(j')}=k^{*(j')}+\alpha_ns^{(j')}/n$ and $0>s^{(j')}\neq\O(1)$ for some index $j'$, and $k^{(j)}=k^{*(j)}+\alpha_ns^{(j)}/n$ and $s^{(j)}=\O(1)$ for $j\neq j'$. Then, we see that $\P\{l(\hat{\bm{\beta}}_{\bm{k}},\bm{k}; \bm{t})-l(\hat{\bm{\beta}}_{\bm{k}^*},\bm{k}^*; \bm{t})>-M\}\to0$ for any $M>0$. Also, through combination with $\eqref{llTaylorBetaHatKStarBetaStarKStar_TIC}$, it follows that $\P\{\hat{ll}(\bm{k};\bm{t},\bm{t})>-M\}\to 0$. As a result, we obtain $K=\{\bm{k}\mid k^{(j)}\in K^{(j)},\ j\in\{1,2,\ldots,m\}\}$, where $K^{(j)}=\{k\mid k-k^{*(j)}=\O(\alpha_n/n)\}$, and \eqref{SupllHat_TIC}, \eqref{ArgsupllHat_TIC}, \eqref{llTaylorBetaStarKTildeBetaHatKTilde_TIC}, \eqref{llTaylorBetaStarKStarBetaStarKTilde_TIC}, and \eqref{llHatKTilde_TIC} thus hold.

Hereafter, by letting $\bm{s}$ be a vector with finite values, we assume that $\bm{k}=\bm{k}^*+\alpha_n\bm{s}/n$. Under the condition in \eqref{condition_TIC}, we have
\begin{align}
\notag 
	&\frac{1}{\sqrt{\alpha_n}} \bm{\Delta}_{\bm{\beta}^{*}}^{(j)\T}
			\Bigg[\sum_{i\in D_{1\alpha_n}^{*(j)}}
			\{\bm{z}_i-\bm{h}(t_i,\bm{\beta}^{*(j)})\}\Bigg] \stackrel{\rm d}{\to}
		(\bm{\Delta}_{\bm{\beta}^{*}}^{(j)\T}\bm{A}_0^{*(j)}\bm{\Delta}_{\bm{\beta}^{*}}^{(j)})^{1/2}W_s
\end{align}
and
\begin{align}
\notag 
	&\frac{1}{\sqrt{\alpha_n}} \bm{\Delta}_{\bm{\beta}^{*}}^{(j)\T}
		\Bigg[\sum_{i\in D_{2\alpha_n}^{*(j)}}
			\{\bm{z}_i-\bm{h}(t_i,\bm{\beta}^{*(j+1)})\}\Bigg] \stackrel{\rm d}{\to}
		(\bm{\Delta}_{\bm{\beta}^{*}}^{(j)\T}\bm{A}_0^{*(j+1)}\bm{\Delta}_{\bm{\beta}^{*}}^{(j)})^{1/2}W_s,
\end{align}
where $\{W_s\}_{s\in\mathbb{R}}$ denotes two-sided standard Brownian motion. Furthermore, in \eqref{C1} and \eqref{C2}, we replace $\bm{\Delta}_{\bm{\beta}_{\xi}^{*}}^{(j)}$, $\bm{\beta}_{\xi}^{*(j)}$, and $\bm{B}_{\xi}^{*(j)}$ with $\bm{\Delta}_{\bm{\beta}^{*}}^{(j)}$, $\bm{\beta}^{*(j)}$, and $\bm{B}_0^{*(j)}$, respectively. Then, by defining $V_{s}^{*(j)}$ as
\begin{align*}
V_s\bigg\{\frac{1}{2}\bm{\Delta}_{\bm{\beta}^*}^{(j)\T}\bm{B}_0^{*(j)}\bm{\Delta}_{\bm{\beta}^{*}}^{(j)},
							\frac{1}{2}\bm{\Delta}_{\bm{\beta}^*}^{(j)\T}\bm{B}_0^{*(j+1)}\bm{\Delta}_{\bm{\beta}^{*}}^{(j)}, 
							(\bm{\Delta}_{\bm{\beta}^{*}}^{(j)\T}\bm{A}_0^{*(j)}\bm{\Delta}_{\bm{\beta}^{*}}^{(j)})^{1/2},
							(\bm{\Delta}_{\bm{\beta}^{*}}^{(j)\T}\bm{A}_0^{*(j+1)}\bm{\Delta}_{\bm{\beta}^{*}}^{(j)})^{1/2}\bigg\},
\end{align*}
it follows that $Q_{k^*+\alpha_ns/n, \bm{t}}^{(j)} \stackrel{\rm d}{\to}V_s^{*(j)}$. As a consequence of this convergence, we obtain
\begin{align}
\notag 
	&\sup_{k\in K^{(j)}}Q_{k,\bm{t}}^{(j)}
		\stackrel{\rm d}{\to}
		\sup_{s\in\mathbb{R}}V_s^{*(j)}
\end{align}
and
\begin{align}
\notag 
	&Q_{\argsup_{k\in K^{(j)}}Q_{k,\bm{u}}^{(j)},\bm{t}}^{(j)}
		\stackrel{\rm d}{\to}
		V_{\argsup_{s\in\mathbb{R}}V'^{*(j)}_s}^{*(j)},
\end{align}
where $V'^{*(j)}_s$ is a copy of $V_s^{*(j)}$. Hence, by the same reasoning as for \eqref{BaBaS} and \eqref{Stryhn}, we can evaluate these expectations and obtain the following corollary.

\begin{corollary}
Under the conditions in Theorem \ref{corollary2} and \eqref{condition_TIC}, even if model misspecification with $\xi=0$ in \eqref{rpll} exists, the asymptotic bias in \eqref{zenbias} is given by
	\begin{align}
		\E\{b(\bm{k}^*,\bm{\beta}^*)\}
			&=2\sum_{j=1}^m C(\bm{A}_0^{*(j)}, \bm{B}_0^{*(j)})+\sum_{j=1}^{m+1}\text{tr}(\bm{A}_0^{*(j)}\bm{B}_0^{*(j)-1}).
\notag 
	\end{align}
	\label{corollary3}
\end{corollary}

\noindent
While this gives an information criterion via the bias-corrected maximum log-partial likelihood, because the asymptotic bias in \eqref{corollary3} contains unknown parameters, they are replaced by consistent estimators, as in \eqref{henkaAICKappa}. As a result, we propose the following information criterion for the Cox proportional hazards model with change-points in cases of model misspecification:
\begin{align}
	\text{TIC}=-2l(\hat{\bm{\beta}},\hat{\bm{k}};\bm{t}) 
		&+4\sum_{j=1}^m \hat{\bm{C}}\{\hat{\bm A}_0^{(j)}(\hat{\bm{\beta}}, 
			\hat{\bm{k}}),\hat{\bm B}_0^{(j)}(\hat{\bm{\beta}}, \hat{\bm{k}})\}
			+2\sum_{j=1}^{m+1}\text{tr}\{\hat{\bm A}_0^{(j)}(\hat{\bm{\beta}}, 
			\hat{\bm{k}})\hat{\bm B}_0^{(j)}(\hat{\bm{\beta}}, \hat{\bm{k}})^{-1}\},
\notag 
\end{align}
where
\begin{align*}
	&\hat{\bm{A}}_0^{(j)}(\hat{\bm{\beta}}, \hat{\bm{k}})
		\equiv \frac{1}{n}\bm{w}^{(j)}(\hat{\bm{\beta}},\hat{\bm{k}})\bm{w}^{(j)}(\hat{\bm{\beta}},\hat{\bm{k}})^\T
\end{align*}
and
\begin{align*}
	&\hat{\bm{B}}_0^{(j)}(\hat{\bm{\beta}}, \hat{\bm{k}})
		\equiv \frac{1}{n}\sum_{i\in \hat{D}^{(j)}}\{\bm{H}(t_i,\hat{\bm{\beta}}^{(j)})
			-\bm{h}(t_i,\hat{\bm{\beta}}^{(j)})\bm{h}(t_i,\hat{\bm{\beta}}^{(j)})^\T\}.
\end{align*}

\section{Conclusion}
\label{sec7}
In light of the high demand for change-point detection with respect to the hazard function in survival time analysis, this paper has derived AIC-type information criteria for the Cox proportional hazards model with change-points for estimation based on the partial likelihood method. First, we evaluated the asymptotic bias of the regularized maximum log-partial likelihood, and we showed via Theorem \ref{theorem1} that the asymptotic bias caused by a change-point can be expressed in terms of the expectation for a two-sided random walk with a negative drift. Then, by assuming an additional natural condition, which is often imposed in asymptotics for change-point analysis, we showed via Theorem \ref{theorem2} that the asymptotic bias can be expressed in a simple, explicit form. As a result, we demonstrated that the AIC can be obtained without any difficulties when estimated by the regularized partial likelihood method, and that the asymptotic bias due to the change-point parameter can be more easily written as 3 when there is no regularization term. The model here has a different aspect from conventional change-point models in that the time and outcome variables are the same. This indicates the need for new asymptotics; however, as long as the partial likelihood method is used, it is sufficient to deal with conventional asymptotics. Indeed, the asymptotic unbiased estimator derived in this paper for the Kullback-Leibler divergence between the true and estimated distributions is similar to that of a conventional change-point model.

Through numerical experiments, we demonstrated that the asymptotic bias evaluated in this paper could be approximated with high accuracy. Furthermore, regarding the original purpose of AIC-type information criteria, which is to give an estimate close to the true structure, the proposed AIC gave clearly smaller Kullback-Leibler divergences than the formal AIC. Moreover, through real data analysis, we indicated that the formal AIC seemed to cause overfitting and it would easily lead to different results from the proposed AIC. Although this paper addressed change-point analysis as a method of mitigating the proportional hazard property in the Cox model, the proposed AIC also relied on log-linearity. Accordingly, we extended it to the TIC that is theoretically guaranteed even when the model is misspecified.

Although we addressed a model in which the hazard function changes with time (i.e., a change-point model for time), models in which the hazard function changes with covariate values (i.e., change-point models for covariates) have also been discussed, especially in recent years (e.g., \citealt{pons2003estimation}, \citealt{lee2020testing}, \citealt{lee2020survival}, and \citealt{wang2021change}). While a change-point model for time basically considers a jump model with abrupt changes, change-point models for covariates also often consider a model with gradual changes. In a jump-type model for covariates, as in a change-point model for time, the estimator of the change-point parameter has been reported to converge faster than that of the regression parameter. In contrast, in a gradual-change model for covariates, the estimator of the change-point parameter converges at the same rate as that of the regression parameter, and it has been reported to have asymptotic normality. For change-point detection, test-based methods using asymptotic normality have been proposed. However, information-criterion-based methods have not been proposed, and it will be necessary to develop them. Because the convergence speeds for the estimators are different in the two change-point models, which implies a difference in the accuracies of the estimators, we expect that the penalty terms of the information criteria for the two models will be considerably different. Specifically, the instances of two-sided Brownian motion appearing in the limit are expected to be different, which will necessitate a new evaluation of the expectation.

In survival time analysis, joint modeling, which simultaneously models repeatedly measured covariates and survival time data, has gained attention (see, e.g., \citealt{HenDD00}). Because this approach is an extension of the Cox proportional hazards model, change-point analysis for both time and covariates in this model will be necessary. The first difficulty is in the use of the profile likelihood, which can be regarded as an extension of the partial likelihood. The asymptotic theory was constructed in \cite{ZenC05}, and the problem will be to tune the theory and reconcile it with the theory used in this paper. Another difficulty in joint modeling is that construction of the information criterion itself is also a hurdle. Regarding this difficulty, because we will deal with a semiparametric model that is also for repeatedly measured covariates and usually includes a random-effects term, the construction of the AIC or conditional AIC, as in \cite{xu2009using} and \cite{DonOXV11}, will not be trivial.

\section*{Acknowledgement}

This research was supported by a JSPS Grant-in-Aid for Scientific Research (16K00050).

\bibliography{List}

\end{document}